\documentclass[12pt]{elsarticle}

\usepackage{amssymb}
\usepackage{amsmath}
\usepackage{graphicx}
\usepackage{bm}
\usepackage{epsfig}
\usepackage{multirow}
\usepackage{lineno}
\usepackage{slashbox}
\usepackage{array}
\usepackage{colortbl}
\usepackage{lscape}

\journal{Nuclear Instruments and Methods A}
\begin{document}
\begin{frontmatter}
\title{Chi-Square Tests  for  Comparing  Weighted Histograms}
\author{N.D. ~Gagunashvili\corref{cor1}\fnref{fn1}}
\ead{nikolai@unak.is}
\cortext[cor1]{Tel.: +3544608505; fax: +3544608998}
\fntext[fn1]{Present address: Max-Planck-Institut f\"{u}r Kernphysik, PO Box
103980, \\ 69029 Heidelberg, Germany}
\address{University of Akureyri, Borgir, v/Nordursl\'od, IS-600 Akureyri, Iceland }
\begin{abstract}
Weighted histograms in Monte Carlo simulations are often used for the estimation of probability density functions. They are obtained as a result of random experiments with random events that have weights. In this paper, the bin contents of a weighted histogram are considered as a sum of random variables with a random number of terms. Generalizations of the classical chi-square test for comparing weighted histograms are proposed. Numerical examples illustrate an application of the tests for the histograms with different statistics of events and different weighted functions. The proposed tests can be used for the comparison of experimental data histograms with simulated data histograms, as well as for the two simulated data histograms.
\end{abstract}
\end{frontmatter}
\section{Introduction}
  A histogram with $m$ bins for a given probability density function  $p(x)$ is used to estimate the probabilities $p_i$ that a random event  belongs in  bin $i$:
\begin{equation}
p_i=\int_{S_i}p(x)dx, \; i=1,\ldots ,m. \label{p1}
\end{equation}
 Integration in (\ref{p1}) is carried out over the bin  $S_i$ and $\sum_1^m p_i=1$.
A histogram can be obtained as a result of a random experiment with the probability
 density function $p(x)$.

A frequently used technique in data analysis is the comparison of two distributions through the comparison of histograms.
The  hypothesis of homogeneity  \cite{cramer} is that the two histograms
  represent random  values with  identical distributions.
  It is equivalent to there existing $m$ constants $p_1,...,p_m$,
 such that $\sum_{i=1}^{m} p_i=1$,
 and the probability  of  belonging  to the  $i$th bin for some  measured value
 in both experiments is  equal to $p_i$.

Let us denote the number of random events belonging to the $i$th bin of the first and second histograms as $n_{1i}$ and $n_{2i}$, respectively. The total number of events in the histograms are equal to $n_j=\sum_{i=1}^{m}{n_{ji}}$, where $j = 1, 2$. Note that over/underflows have to taken into account for
  these relations to hold.

It has been shown in Ref. \cite{pearson} that the statistic
\begin{equation}
\sum_{i=1}^{m} \frac{(n_{ji}-n_jp_{i})^2}{n_jp_{i}} \label{basic}
\end{equation}
 has  approximately a $\chi^2_{m-1}$ distribution.
For two statistically independent histograms with probabilities $p_1,...,p_m$  the statistic
\begin{equation}
\sum_{j=1}^2  \sum_{i=1}^{m} \frac{(n_{ji}-n_jp_{i})^2}{n_j p_{i}} \label{basic2}
\end{equation}
has approximately a $\chi^2_{2m-2}$ distribution.
 If the probabilities $p_1,...,p_m$ are not known, then they can be estimated by the minimization of Eq. (\ref{basic2}). The estimation of $p_i$ is carried out by the following expression:
 \begin{equation}
 \hat{p}_i= \frac{n_{1i}+n_{2i}}{n_{1}+n_{2}} \, \text{ ,} \label{phat1}
 \end{equation}
 as shown in Ref. \cite{cramer}.
By substituting expression (\ref{phat1}) in Eq. (\ref{basic2}), the statistic
\begin{equation}
\sum_{j=1}^2  \sum_{i=1}^{m} \frac{(n_{ji}-n_j\hat{p}_{i})^2}{n_j \hat{p}_{i}}
=\frac{1}{n_1n_2} \sum_{i=1}^{m}{\frac{(n_2n_{1i}-n_1n_{2i})^2}{n_{1i}+n_{2i}}} \label{xsquar1}
\end{equation}
is obtained.
This statistic has approximately a $\chi^2_{m-1}$ distribution
 because  $m-1$ parameters  are  estimated \cite{cramer}.
 The statistic  (\ref {xsquar1}) was first  developed in \cite{fisher} and is widely used to test the  hypothesis of homogeneity.


Weighted histograms are often obtained as a result of Monte-Carlo
simulations.
References \cite{muon,weight,astro} are examples of research on high-energy physics, statistical mechanics, and astrophysics using such histograms.
 Operations with weighted
histograms have been  realized in contemporary systems for data analysis
HBOOK \cite{hbook}, Physics Analysis Workstation (PAW) \cite{paw} and the
ROOT framework \cite{root1}, developed at CERN (European Organization
for Nuclear Research, Geneva, Switzerland).

To define a weighted histogram let us write the probability $p_i$
(\ref{p1}) for a given probability density function  $p(x)$  in the
form
\begin{equation}
p_i= \int_{S_i}p(x)dx = \int_{S_i}w(x)g(x)dx, \label{weightg}
\end{equation}
where
\begin{equation}
w(x)=p(x)/g(x) \label{fweight}
\end{equation}
 is the weight function and $g(x)$ is some other probability density function. The function $g(x)$ must be $>0$ for points $x$, where $p(x)\neq 0$. The weight $w(x)=0$ if $p(x)=0$, see Ref.  \cite{Sobol}.  Because of the condition $\sum_ip_i=1$ further we will call the above defined weights  normalized weights as opposite to the unnormalized weights $\check{w}(x)$ which are $\check{w}(x)=const\cdot w(x)$.

The histogram with normalized weights
 was obtained from a random experiment with a probability density function $g(x)$, and the weights of the events were calculated according to (\ref{fweight}). Let us denote the total sum of the weights of the events in the $i$th bin of the histogram with normalized weights as
\begin{equation}
W_i= \sum_{k=1}^{n_i}w_i(k), \label{ffweight}
\end{equation}
where $n_i$ is the number of events at bin $i$ and $w_i(k)$ is the weight of the $k$th event in the $i$th bin. The total number of events in the histogram is equal to $n=\sum_{i=1}^{m}{n_i}$, where $m$ is the number of bins. The quantity $\hat{p}_i= W_{i}/n$ is the estimator of $p_i$ with the expectation value $\textrm E \,
\hat{p_i}=p_i$. Note that in the case where $g(x)=p(x)$, the weights of the events are equal to 1 and the histogram with normalized weights
is the usual histogram with unweighted entries.

 Nowadays, the apparatus used for measurements have become more complex and computers have become more powerful. The final theoretical prediction of a model is often obtained by Monte Carlo simulation and often with the usage of weights for the simulated events.

  Comparison of two weighted histograms, comparison of the weighted histogram and the histogram with unweighted entries, as well as the fitting weights of simulated random events to the experimental histogram are all important parts of data analysis.

 The problem of fitting experimental histograms using simulated model histograms (weighted histograms) has been discussed in Ref. \cite{schmidt} . Information about the statistical uncertainties of the weighted histograms is not used in the fitting algorithm, and the method proposed in Ref. \cite{schmidt} can be recommended for use with very high statistics of Monte Carlo simulations.

 Another method that takes into account the statistical errors of simulated theoretical prediction was proposed in
  Ref. \cite{eberhard} for the special case of linear superposition of several model distributions produced by a
  parameter-free Monte Carlo simulation.

   On the other hand, a common approach for the comparison of the weighted
  histogram and histograms with unweighted entries was developed in Ref. \cite{kortner}. Unfortunately, the formula (32) for the chi-square
   test generalization on page 633 of Ref. \cite{kortner} cannot be used for the cases where the histograms have
    different total number of events. To prove this statement, it is sufficient to consider the formula for the
     case of the two histograms with unweighted entries. The formula coincides with statistic (\ref {xsquar1}) for the case of
      the two histograms with equal total number of events, and does not lead to statistic (\ref {xsquar1})
       when the number of events is different. In the same way, it is not difficult to prove that all other
        formulas presented Ref. \cite{kortner} cannot be used for the comparison of histograms with different
         total number of events; this is a serious restriction with respect to the practical application of
          the proposed approach.

Modified chi-square tests for the comparison of the weighted histograms and histograms with unweighted entries were proposed
in Ref. \cite{gagunashvili,gagunashvili1}. The proposed tests are available in the ROOT framework under
the class TH1:Chi2Test \cite{root}. The main disadvantage of these tests is the rather high minimal number
 of events for the bins of the weighted histogram, which is equal to 25. In addition, the tests do not work
 properly if the total number of events for one histogram is considerably greater than that for another.

Among the approaches widely used in practice, the heuristic chi-square test presented in Ref. \cite{zech}
 is well known. To test the hypothesis of homogeneity, the author had proposed to normalize the histograms
  with respect to each other and use the statistics
 \begin{equation}
 X_h^2=\sum_{i=1}^m \frac{(W_{1i}-W_{2i})^2}{d^2_i}, \label{heurist}
 \end{equation}
where $W_{ji},j=1,2; i=1,...,m$ is the  sum of weights in $i$th
bin of $j$th histogram and
\begin{gather}
d_i^2= \begin{cases}
       W_{2i}\, [s^2(W_{1i})/W_{1i}+ s^2(W_{2i})/W_{2i}] & n_2 \geq n_1\\
       W_{1i}\, [s^2(W_{1i})/W_{1i}+ s^2(W_{2i})/W_{2i}] & n_2 < n_1
       \end{cases}
\end{gather}
where $s^2(W_{ji})$ is the total sum of the squares of the weights in the $i$th bin of the $j$th histogram.
The number of events per bin should be $>20$. It is expected that if the hypothesis of homogeneity is true,
 then the statistics $X_h^2$  has a chi-square distribution; however, the number of degrees of freedom was
 not clearly defined in Ref. \cite{zech}. Note that statistic (\ref{heurist}) for the case of the
 two histograms with unweighted entries does not lead to the classical chi-square statistic (\ref{xsquar1}).

 Recently, a goodness-of-fit test for the weighted histograms was proposed in Ref.  \cite{gagunashvili3}.
 This test is a generalization of Pearson's test for weighted histograms and leads to the usual Pearson's
 test for the case of histograms with unweighted entries.
In this paper, we have used these results and have developed the generalization of the chi-square
homogeneity test that, for the case wherein the weights of the events in both the histograms are equal
 to 1, leads to the usual chi-square test. In addition, important practical tests for histograms with unnormalized weights
   have been developed. It has been shown that this new approach presented
  permits an essential decrease in the minimal number of events per bin required when compared with
   Refs. \cite{gagunashvili1, zech}, and can be applied for the case of different total number
    of events in the histograms.

The paper is organized as follows. In Section 2, a generalization of the chi-square homogeneity test
 is proposed. The test for the histograms with unnormalized weights is discussed in Section 3.
  Application and verification of the tests are demonstrated using numerical examples in Section 4.
   Furthermore, the sizes of the tests are compared with the calculated sizes of the heuristic
    chi-square test \cite{zech} for important practical case of  histogram with unweighted entries (experimental data histogram)
     and histogram with unnormalized weights (simulated data histogram). The comparison has demonstrates the superiority
      of the proposed generalization of the chi-square test over the heuristic chi-square test \cite{zech} .

\section{ Homogeneity test for comparison two histograms with normalized weights}

Let us consider two  histograms with normalized weights, and subindex $j$ will be used to differentiate them. The total
 sum of weights of events $W_{ji}$ in the $i$th bin of the $j$th histogram $j=1,2$;  $i=1,\ldots,m$ can
  be considered as a sum of random variables
 \begin{equation}
W_{ji}= \sum_{k=1}^{n_{ji}}w_{ji}(k), \label{ffffweight}
\end{equation}
where the number of events  $n_{ji}$ is also a random value and the
weights $w_{ji}(k),k=1,...,n_{ji}$ are independent
 random variables with the same probability distribution function for given bin.
 Let us introduce a variable
\begin{equation}
r_{ji}=\textrm E \, w_{ji}/\textrm E \, w_{ji}^2 \label{rati}
\end{equation}
 which is the ratio of the first moment  to the second moment of the distribution of weights in bin $i$.
As shown in Ref. \cite{gagunashvili3} the statistic
\begin{equation}
 \frac{1}{n_j} \sum_{i \neq k} \frac{r_{ji}W_{ji}^2}{p_{i}}+\frac{1}{n_j}
\frac{(n_j-\sum_{i \neq k}r_{ji}W_{ji})^2}{1-\sum_{i \neq k}r_{ji}
p_{i}}-n_j\label{stddu}
\end{equation}
 where sums extends over all bins $i$ except one bin $k$,
 approximately has a $\chi^2_{m-1}$ distribution and is a generalization of the Pearson's statistic (\ref{basic}).

 Note that denominator $1-\sum_{i \neq k}r_{ji}p_{i} > 0$. To prove this statement let us write  the probability $p_{i}$ as
 \begin{equation}
 p_{i}=g_{ji}\textrm E \, w_{ji}
 \end{equation}
 then
 \begin{equation}
\sum_{i \neq k}r_{ji}p_{i}=\sum_{i \neq k}g_{ji}\frac{(\textrm E \, w_{ji})^2}{\textrm E \, w_{ji}^2} \leq \sum_{i \neq k}g_{ji} < 1
\end{equation}
because following  H\"{o}lder's  inequality
\begin{equation}
(\textrm E \, w_{ji})^2 \leq \textrm E \, w_{ji}^2.
\end{equation}

For two statistically independent histograms with probabilities $p_1,...,p_m$  the statistic
\begin{equation}
X_k^2=\sum_{j=1}^2\frac{1}{n_j} \sum_{i \neq k} \frac{r_{ji}W_{ji}^2}{p_{i}}+\sum_{j=1}^2\frac{1}{n_j}
\frac{(n_j-\sum_{i \neq k}r_{ji}W_{ji})^2}{1-\sum_{i \neq k}r_{ji}
p_{i}}-\sum_{j=1}^2n_j\label{stdd2}
\end{equation}
has approximately a $\chi^2_{2m-2}$ distribution.
The probabilities, $p_i$ , are not known and $\hat p_1,\ldots,\hat p_{k-1},\hat p_{k+1},\ldots,\hat p_m$ can be determined by the minimization of Eq. (\ref{stdd2}) under constraints
 \begin{equation}
  \hat p_i > 0,\,\,  1-\sum_{i \neq k}\hat p_i > 0,\,\, 1-\sum_{i \neq k}r_{1i}\hat p_i > 0, \text { and } 1-\sum_{i \neq k}r_{2i} \hat p_i>0\label{constr}.
 \end{equation}
Subsequently, the homogeneity test statistic will have a $\chi^2_{m-1}$ distribution, because $m-1$ parameters are estimated \cite {fisher}.
Let us now replace $r_{ji}$ with the estimate
\begin{equation}
\hat r_{ji}=\sum_{k=1}^{n_{ji}}w_{ji}(k)/\sum_{k=1}^{n_{ji}}w_{ji}^2(k).\label{rat4}
\end{equation}
Then, the test statistic is given as
\begin{equation}
\hat X_k^2=\sum_{j=1}^2\frac{1}{n_j} \sum_{i \neq k} \frac{\hat r_{ji}W_{ji}^2}{\hat p_{i}}+\sum_{j=1}^2\frac{1}{n_j}
\frac{(n_j-\sum_{i \neq k}\hat r_{ji}W_{ji})^2}{1-\sum_{i \neq k}\hat r_{ji}
\hat p_{i}}-\sum_{j=1}^2n_j .\label{stdd3}
\end{equation}

Note that for the histograms with unweighted entries $W_{ji}=n_{ji}$, $\hat p_i= (n_{1i}+n_{2i})/(n_1+n_2)$, $\hat r_{ji}=1$, statistic (\ref{stdd3}) coincides with the chi-square statistic (\ref{xsquar1}). Although the estimators of the probabilities (\ref{phat1}) for the histograms with unweighted entries were found, the common problem of minimization of Eq. (\ref{stdd3}) to determine the estimators of probabilities $\hat p_i$ cannot be effectively solved analytically. However, the problem has been solved numerically by the coordinate-wise optimization in this paper. For each step, the minimum is found for one probability with the others fixed, using the Brent algorithm \cite{brent, cern}. The rather good initial approximation
\begin{equation}
\hat p_i=\frac {\hat r_{1i}W_{1i}+\hat r_{2i}W_{2i}}{\hat r_{1i}n_{1}+\hat r_{2i}n_{2}}
\end{equation}
provides a fast convergence of the algorithm to the minimum with an easy control under constraints
 (\ref {constr}). Formula (\ref{stdd3}) for the histograms with unweighted entries does not depend on the choice of
  the excluded bin; however, for the histograms with normalized weights, there can be a dependence. A test statistic
   that is invariant to the choice of the excluded bin can be obtained as the median value of formula
    (\ref{stdd3}) for a different choice of the excluded bin
\begin{equation}
\hat X^2= \textrm {Med }\, \{\hat X_1^2,  \hat X_2^2,  \ldots , \hat
X_m^2\}.\label{stdavu}
\end{equation}
as carried out in Ref.  \cite{gagunashvili3}, for the goodness-of-fit test.

The chi-square approximation is asymptotic. This means that the critical values may not be valid if the expected frequencies are too small. The use of the chi-square test is inappropriate
 if any expected frequency is $<1$, or if the expected frequency is $<5$ in $>20\%$ of the bins
  \cite{moore,cochran} for either histogram. This restriction observed in the usual chi-square test is quite reasonable for the proposed test.

Note  that for the case $W_{ji}=0$,  the ratio $\hat r_i$ is
undefined. The average value of this quantity for its nearest neighbors bins with non-zero
 bin content can be used for an approximation of the undefined  $\hat r_i$ or the  empty bin can be merged with the nearest neighbor
  bin that is not empty. Moreover, the last possibility is $\hat r_i=0$ that makes the test more conservative.

\section{The test for histograms with unnormalized weights}

In practice one is often confronted with the case when a histogram is defined up to an unknown normalization constant. Let us denote a bin
content of histograms with unnormalized weights as $\check{W_{ji}}$, then
$W_{ji}=\check{W_{ji}}C_j$, and the test statistic (\ref{stddu}) can be
written as
\begin{equation}
 \frac{C_j}{n_j} \sum_{i \neq k} \frac{\check{r}_i\check{W}_{ji}^2}{p_{i}}+\frac{1}{n_j}
\frac{(n_j-\sum_{i \neq k}\check{r}_{ji}\check{W}_{ji})^2}{1-C_j^{-1}\sum_{i
\neq k}\check{r}_{ji} p_{i}}-n_j,\label{stddc}
\end{equation}
with $\check{r}_{ji}=C_jr_{ji}$.
An estimator $\hat C_{kj}$  for the constant $C_j$ \ is
found  in \cite{gagunashvili3} by minimization of Eq. (\ref{stddc}) and equal to
\begin{equation}
\hat C_{kj}=\sum_{i \neq k}\check{r}_{ji}p_{i}+\sqrt{\frac{\sum_{i
\neq k}\check{r}_{ji}p_{i}}{\sum_{i \neq
k}\check{r}_{ji}\check{W}_{ji}^2/p_{i}}}(n_j-\sum_{i \neq
k}\check{r}_{ji}\check{W}_{ji}). \label{const}
\end{equation}
 Substituting (\ref{const}) to the (\ref{stddc}) and replacing $\check{r}_{ji}$ with the estimate  $\hat{\check{
r}}_{ji}$  we get the test statistic
\begin{equation}
\frac{s_{kj}^2}{n_j}+2s_{kj}, \label{sss69}
\end{equation}
where
\begin{equation}
s_{kj}=\sqrt{\sum_{i \neq k}\hat{\check{r}}_{ji} p_{i} \sum_{i \neq k}
\hat{\check{r}}_{ji}{\check{W}}_{ji}^2/p_{i}} - \sum_{i \neq
 k}\hat{\check{r}}_{ji}{\check{W}}_{ji}.\label{statnorm}
\end{equation}
The estimate $\hat{\check{r}}_{ji}$ in (\ref{statnorm}) is calculated in the same way as the estimate $\hat{r}_{ji}$, see formula (\ref{rat4}).

Statistic (\ref{sss69}) has an approximately $\chi^2_{m-2}$ distribution.
  Note that this test, as is shown in the numerical examples in Ref. \cite{gagunashvili3},
  is liberal; in other words, the real size of the test is slightly larger than the nominal value of the
   test. For two statistically independent histograms with probabilities $p_1,..., p_m$, the statistic
\begin{equation}
\check{X}_k^2=\sum_{j=1}^2\frac{s_{kj}^2}{n_j}+2\sum_{j=1}^2 s_{kj},\label{stattwo}
\end{equation}
has approximately a $\chi^2_{2m-3}$ distribution. One degree of freedom is lost because,
as mentioned earlier, the goodness-of-fit tests (\ref{sss69}) are liberal and this effect
accumulates for the two histograms. The probabilities, $p_i$ , are not known and the estimators
 $\hat p_1,\ldots,\hat p_{k-1},\hat p_{k+1},\ldots,\hat p_m$  can be found by the minimization of Eq. (\ref{stattwo}).
Subsequently, the test statistic
 \begin{equation}
\hat {\check{X}}_k^2= \sum_{j=1}^2\frac{\hat s_{kj}^2}{n_j}+2\sum_{j=1}^2 \hat s_{kj},\label{stattwo1}
\end{equation}
 where
 \begin{equation}
 \hat s_{kj}=\sqrt{\sum_{i \neq k}\hat{\check{r}}_{ji}\hat p_{i} \sum_{i \neq k}
\hat{\check{r}}_{ji}\check{W}_{ji}^2/\hat p_{i}} - \sum_{i \neq
 k} \hat{\check {r}}_{ji}\check{W}_{ji}
 \end{equation}
 has $\chi^2_{m-2}$  distribution.
 The probabilities $\hat p_{i}$ can be calculated in the same way as described in Section 2 with the initial approximation
\begin{equation}
\hat p_i=\frac {\hat{\check{r}}_{1i}\check{W}_{1i}+\hat {\check{r}}_{2i}\check{W}_{2i}}{\hat{\check{r}}_{1i}\sum_{i=1}^m \check{W}_{1i}+\hat {\check{r}}_{2i}\sum_{i=1}^m \check{W}_{2i} }.
\end{equation}
A test statistic that is ``invariant" to the choice of the excluded bin can be obtained again as the
 median value of (\ref{stattwo}) for all possible choices of the excluded bin
\begin{equation}
\hat {\check{X}}^2= \textrm {Med }\, \{\hat {\check{X}}_1^2, \hat
{\check{X}}_2^2, \ldots , \hat {\check{X}}_m^2\}.\label{stdav}
\end{equation}
 As a result of the calculations presented in Section 2 and the above-obtained results, the test
 statistics for the comparison of the histogram with normalized
 weights and that with unnormalised weights, can be given as
 \begin{equation}
 \hat {\tilde{X}}_k^2=\frac{1}{n_1} \sum_{i \neq k} \frac{\hat r_{1i}W_{1i}^2}{\hat p_{i}}+\frac{1}{n_1}
\frac{(n_1-\sum_{i \neq k}\hat r_{1i}W_{1i})^2}{1-\sum_{i \neq k}\hat r_{1i}
\hat p_{i}}-n_1+ \frac{\hat s_{k}^2}{n_2}+2\hat s_{k}, \label{sss}
\end{equation}
where
\begin{equation}
\hat s_{k}=\sqrt{\sum_{i \neq k}\hat{\check{r}}_{2i} \hat p_{i} \sum_{i \neq k}
\hat{\check{r}}_{2i}\check{W}_{2i}^2/\hat p_{i}} - \sum_{i \neq
 k} \hat{\check{r}}_{2i} \check{W}_{2i} \label{statnormm}
\end{equation}
has approximately  $\chi^2_{m-2}$ distribution.
The probabilities $\hat p_{i}$ can be calculated in the same way as described in Section 2, with the initial approximation
\begin{equation}
\hat p_i=\frac {\hat r_{1i}W_{1i}+\hat{\check{r}}_{2i}\check{W}_{2i}}{\hat{{r}}_{1i}n_{1}+\hat{\check{r}}_{2i}\sum_{i=1}^m\check{W}_{2i} }.
\end{equation}
Statistic (\ref{sss}) for the very important case of comparing the  histogram with unweighted entries and the
 histogram with unnormalized weights can be given as
 \begin{equation}
 \hat {\tilde{X}}_k^2=\frac{1}{n_1} \sum_{i=1}^m \frac{n_{1i}^2}{\hat p_{i}}-n_1+ \frac{\hat s_{k}^2}{n_2}+2\hat s_{k}. \label{sssu}
\end{equation}
Statistic (\ref {sssu}) can be used to compare the experimental and Monte Carlo histograms, as well as
 for the purpose of fitting the Monte Carlo distribution to the data.

\section{Evaluation of the tests' sizes and power}

The hypothesis  of homogeneity $H_0$ is rejected if the test statistic $\hat X^2$ is larger
 than some threshold. The threshold $k_{\alpha }$ for a given nominal
size of test $\alpha$ can be  defined from the equation
\begin{equation}
\alpha = P\,(\chi^2_l>k_{\alpha})=\int_{k_{\alpha}}^{+\infty}
\frac{x^{l/2-1} e^{-x/2}}{2^{l/2}\Gamma(l/2)}dx, \label {kalfa}
\end{equation}
where $l=m-1$.

Let us define the test size $\alpha_s$ for a given nominal test
size $\alpha$ as the probability
\begin{equation}
\alpha_s =  P\,(\hat X^2>k_{\alpha}|H_0).\label {alfas}
\end{equation}
This is the probability that the hypothesis $H_0$ will be rejected if the distribution of the
weights $W_{ji}$,  $j=1,2;\,\, i=1,...,m$, for the bins of the histograms satisfies the
hypothesis $H_0$. The deviation of the test size from the nominal test size is an important test characteristic.

A second important characteristic of the test is the power $\beta$
\begin{equation}
\beta = P\,(\hat X^2 >k_{\alpha}|H_a). \label {beta}
\end{equation}
This is the probability that the hypothesis of homogeneity $H_0$ will be rejected if the distributions
 of the weights $W_{ji}$, $j=1,2$; $i=1,...,m$, of the compared histograms do not satisfy the hypothesis $H_0$.
The same definitions with $l=m-2$ in formula (\ref {kalfa}) can be used for the test statistic $\hat{\check{X}}^2$.

The following is a numerical example to evaluate the sizes and power of the tests. Let us take a distribution
\begin{equation}
p(x)\propto \frac{2}{(x-10)^2+1}+\frac{1}{(x-14)^2+1} \label{weight}
\end{equation}
 defined on the interval $[4,16]$ and representing two so-called Breit-Wigner
  peaks \cite{breit}.
  Three cases of the probability density function $g(x)$ are
  considered (see Fig. 1)

\begin{equation}
g_1(x)=p(x)   \label{prc}
\end{equation}

\begin{equation}
g_2(x)=1/12  \label{flat}
\end{equation}

\begin{equation}
g_3(x)\propto\frac{2}{(x-9)^2+1}+\frac{2}{(x-15)^2+1} \label{real}
\end{equation}

Distribution $g_1(x)$ (\ref{prc}) results in  an  histogram with unweighted entries, while distribution $g_2(x)$ (\ref{flat}) is a uniform distribution
 on the interval [4, 16]. Distribution $g_3(x)$ (\ref{real}) has the same form of parametrization as $p(x)$ (\ref{weight}), but with different values of the parameters.

The sizes of the tests for histograms with the number of bins equal to 5 and 20, with different weighted
 functions, were calculated for a nominal value of size $\alpha$ equal to 0.05. Calculations of test
  sizes $\alpha_s$   were carried out using the Monte Carlo method based on 10000 runs. The results of
   the calculation for the two histograms with normalized weights, two histograms with unnormalized weights, as
    well as histogram with normalized weights and with unnormalized weights one, are presented in Figs. 2-4. Each plot
     contains 9 subplots, with 3 superrows of subplots and 3 supercolumns of subplots. A subplot
      represents test sizes for a pair of statistically independent histograms with weight functions
       $p(x)/g_i(x)$,  $p(x)/g_j(x)$, and different total number
        of events. For example, second superrow of subplots presents the sizes for the pairs of histograms
         with weight functions  $p(x)/g_1(x)$, $p(x)/g_2(x)$; $p(x)/g_2(x)$, $p(x)/g_2(x)$ and $p(x)/g_3(x)$, $p(x)/g_2(x)$.
          To make comparison easier, all the plots have the same scale (intensity of gray color).

Calculations of test sizes $\alpha_s$   were carried out using the Monte Carlo method, therefore it is reasonable to test the hypothesis $H_0^{(1)}:\alpha_s=0.05$ against the
  alternative $H_a^{(1)}: \alpha_s \neq 0.05$. For this purpose $z$ statistics can be used \cite{moore}
\begin{equation}
z=(\hat{\alpha}_s-0.05)/\sqrt{\frac{0.05\times(1-0.05)}{10000}}
\end{equation}
where $\hat{\alpha}_s$ is the calculated value of $\alpha_s$.
If the null hypothesis is true then this test statistic has a standard normal distribution with mean value  0 and standard deviation 1. For the standard normal distribution, 2.5\% of the values lie below the critical value –1.959964, and 2.5\% lie above 1.959964. Therefore, if we are conducting a 2-sided hypothesis test at the 0.05 level of significance, we will accept $H_0^{(1)}$ when $|z| \leq 1.959964$ or  $0.045728\leq\hat{\alpha}_s\leq 0.054272$.
The $``\bullet"$ markers on the plots show regions satisfying hypothesis $H_0^{(1)}$.

Following Ref. \cite {cochran}, a disturbance is regarded as unimportant when $\alpha=0.05$ and the size of test $\alpha_s$ lies between 0.04 and 0.06. In this study, we defined  the size of the test as close to the nominal value if it satisfied the above-mentioned criteria.
  We have only an estimation of $\alpha_s$ therefore it is reasonable to test the  hypothesis  $H_0^{(2)}:0.04 \leq\alpha_s \leq 0.06$ against alternative $H_a^{(2)}: \alpha_s > 0.06$ or $\alpha_s < 0.04$. According to Ref. \cite{kendall} the critical region for $\hat{\alpha}_s$ to test hypothesis $H_0^{(2)}$ at the 0.05 level of significance is the interval with endpoints $x_1$ and $x_2$ that satisfy the system of equations:
 \begin{equation}
\begin{split}
\int \limits_{\frac{x_1-0.04}{\sqrt{\frac{0.04\times 0.96}{10000}}}}
^{\frac{x_2-0.04}{\sqrt{\frac{0.04\times 0.96}{10000}}}}\phi(x)dx=0.95;
\,\,\,\,\,\,\,\,\, \int \limits_{\frac{x_1-0.06}{\sqrt{\frac{0.06\times 0.94}{10000}}}}
^{\frac{x_2-0.06}{\sqrt{\frac{0.06\times 0.94}{10000}}}}\phi(x)dx=0.95 \label{sys}
\end{split}
 \end{equation}
where $\phi(x)$ is probability density function of the standard normal distribution. System (\ref{sys}) has been solved numerically to give us  $x_1= 0.036777$ and $x_2= 0.063906$ therefore  we accept $H_0^{(2)}$ when  $0.036777\leq\hat{\alpha}_s\leq 0.063906 $.
The $``\Box"$ markers on the plots show regions which do not satisfy  hypothesis $H_0^{(2)}$.

Figures 2-4 demonstrate that the test sizes are close to their nominal values for the appropriate number of events in the bins of both the histograms. Moreover, they are reasonably close to the nominal values
  if only one histogram has the appropriate number of events. Tests are conservative when both the
   histograms have inappropriate number of events. Markers ``$\circ$" on the plots show regions with
    an inappropriate number of events, at least in one histogram.
Tables 1 present the sizes of the test for the comparison of the histogram with unweighted entries and the  histogram with unnormalized weights (\ref{sssu}).
These table correspond to the first supercolumn of the plots presented in Figs. 4(a) and 4(b). Table 2
 present the sizes of the heuristic test for the same case of the  histogram with unweighted entries along with histogram with unnormalized weights (\ref{heurist}).
 The degree of freedom of the chi-square distribution has not been clearly
   defined in Ref. \cite{zech}, and we chose $m-1$ that gave the best results. Comparisons
    of Table 1 with  Table 2  showed the superiority of the proposed
     test over the heuristic chi-square test.

\begin{figure}
\centering \vspace*{-1.9 cm} \hspace*{-0.7 cm}
\includegraphics[width=1.15 \textwidth]{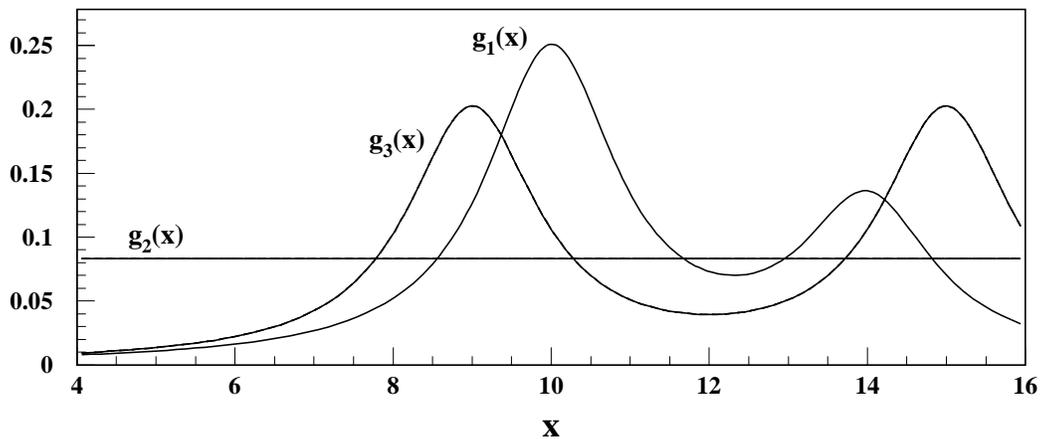}\\
\vspace*{-8. cm}
  \caption {Probability density functions $g_1(x)=p(x)$, $g_2(x)$ and $g_3(x)$.}
\vspace*{1 cm }
\end{figure}
\begin{landscape}
\begin{figure}[h]
\begin{center}$
\begin{array}{ccc}
\includegraphics[width=4.5in]{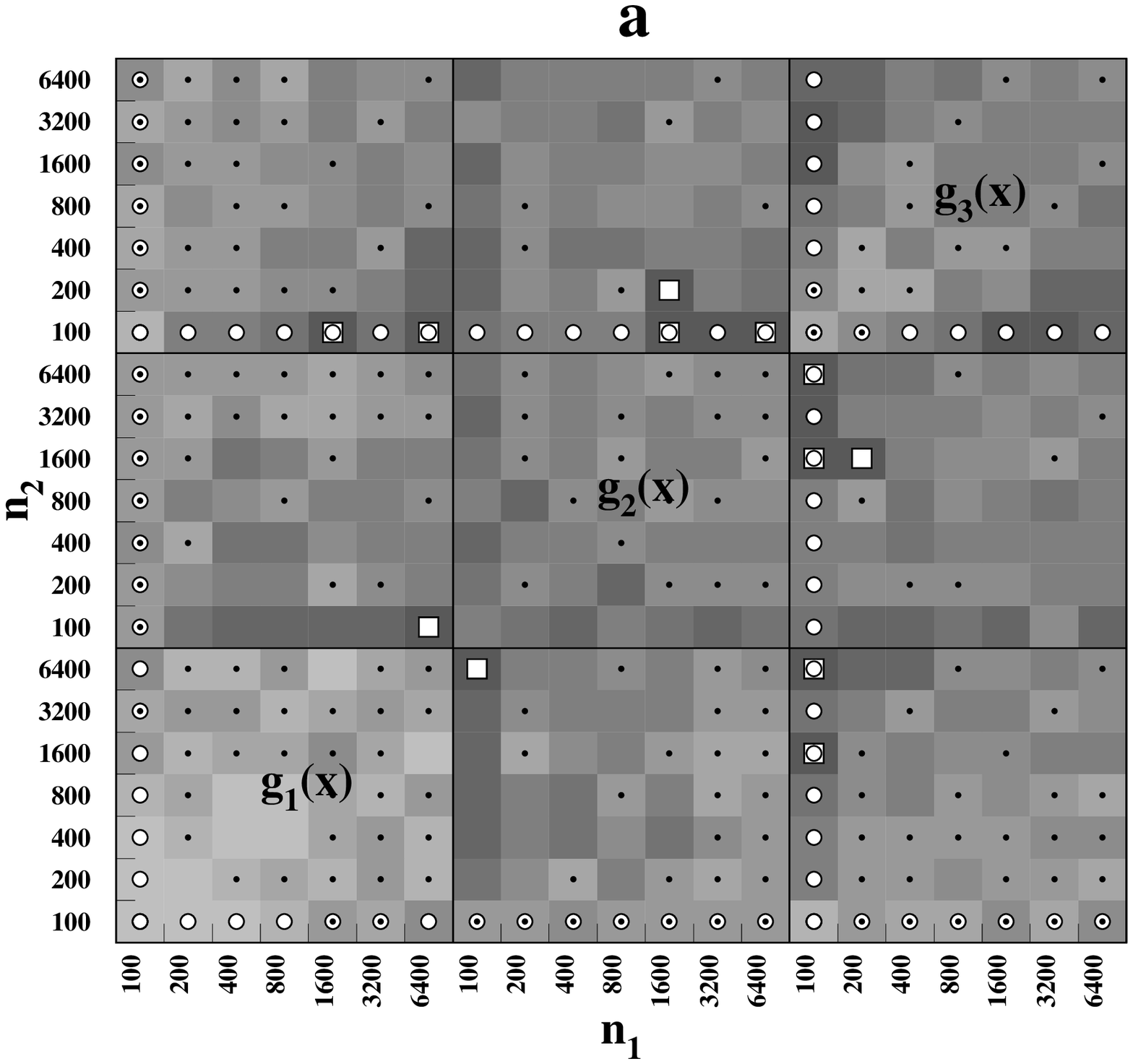}
\hspace*{-2.2 cm}  \includegraphics[width=4.5in]{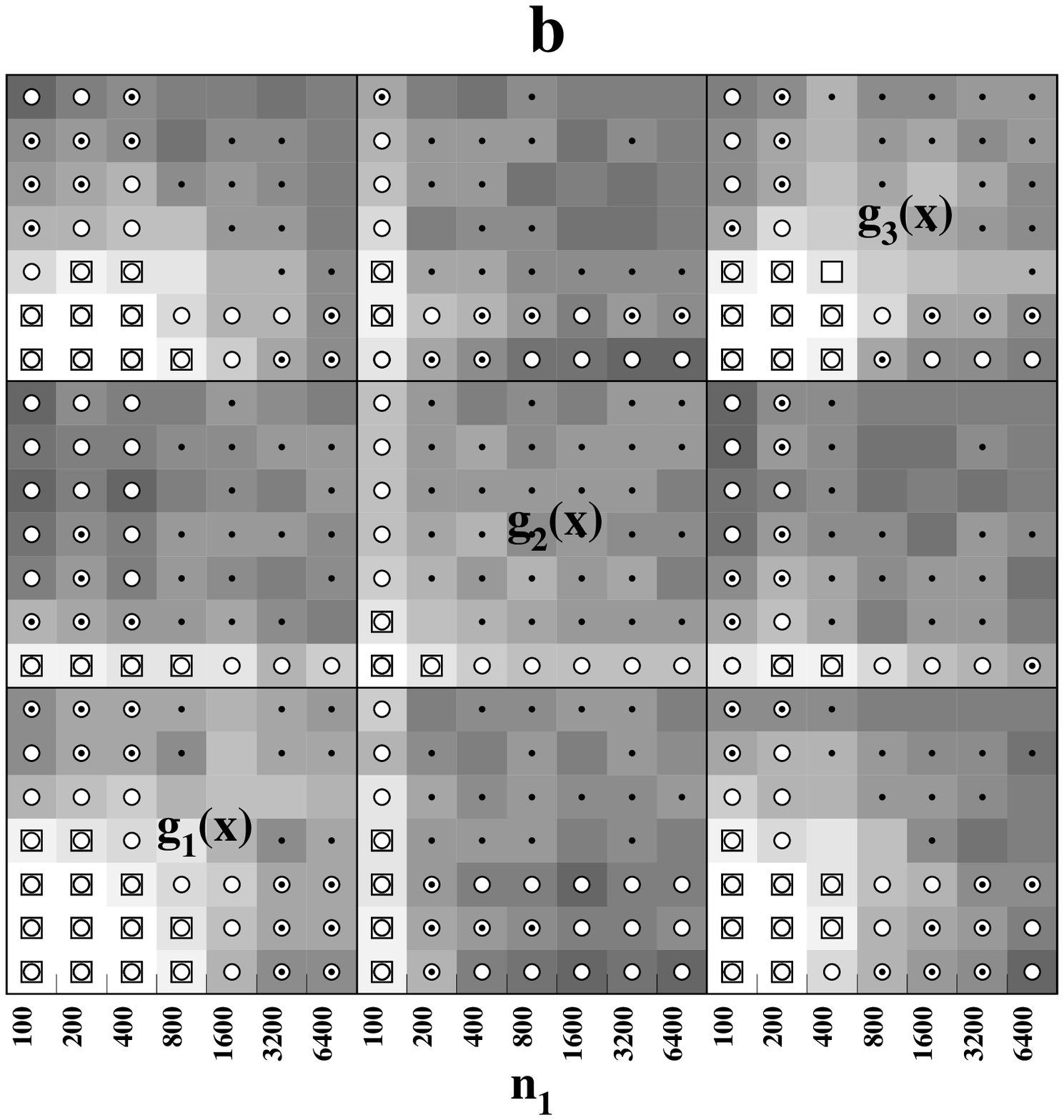} & \hspace*{-2.4 cm}\includegraphics[width=4.5in]{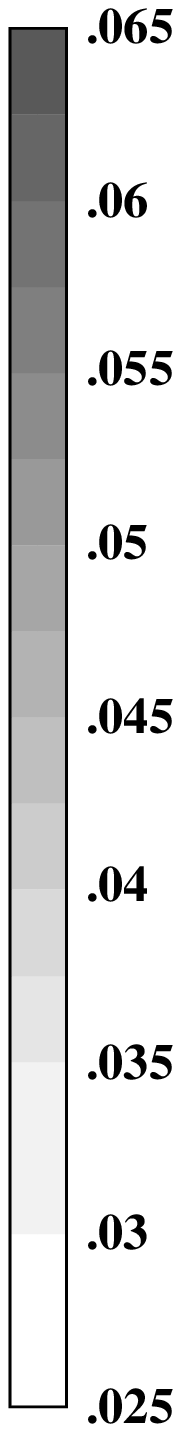}
\end{array}$
\end{center}
\vspace *{-1cm}
\hspace{1cm}\parbox{19cm}{\caption  {Sizes of the test for the comparison of two histograms with normalized weights: (a) number of bins $m=5$, (b) number of bins $m=20$. Markers show regions: ``$\circ$" have inappropriate number of events in the histograms for application of the test, $``\bullet"$ satisfy hypothesis $\alpha_s=0.05$, $``\Box"$ have a size of test $\alpha_s$ that is not close to the nominal size of the test.}}
\end{figure}
\end{landscape}

\begin{landscape}
\begin{figure}[h]
\begin{center}$
\begin{array}{ccc}
\includegraphics[width=4.5in]{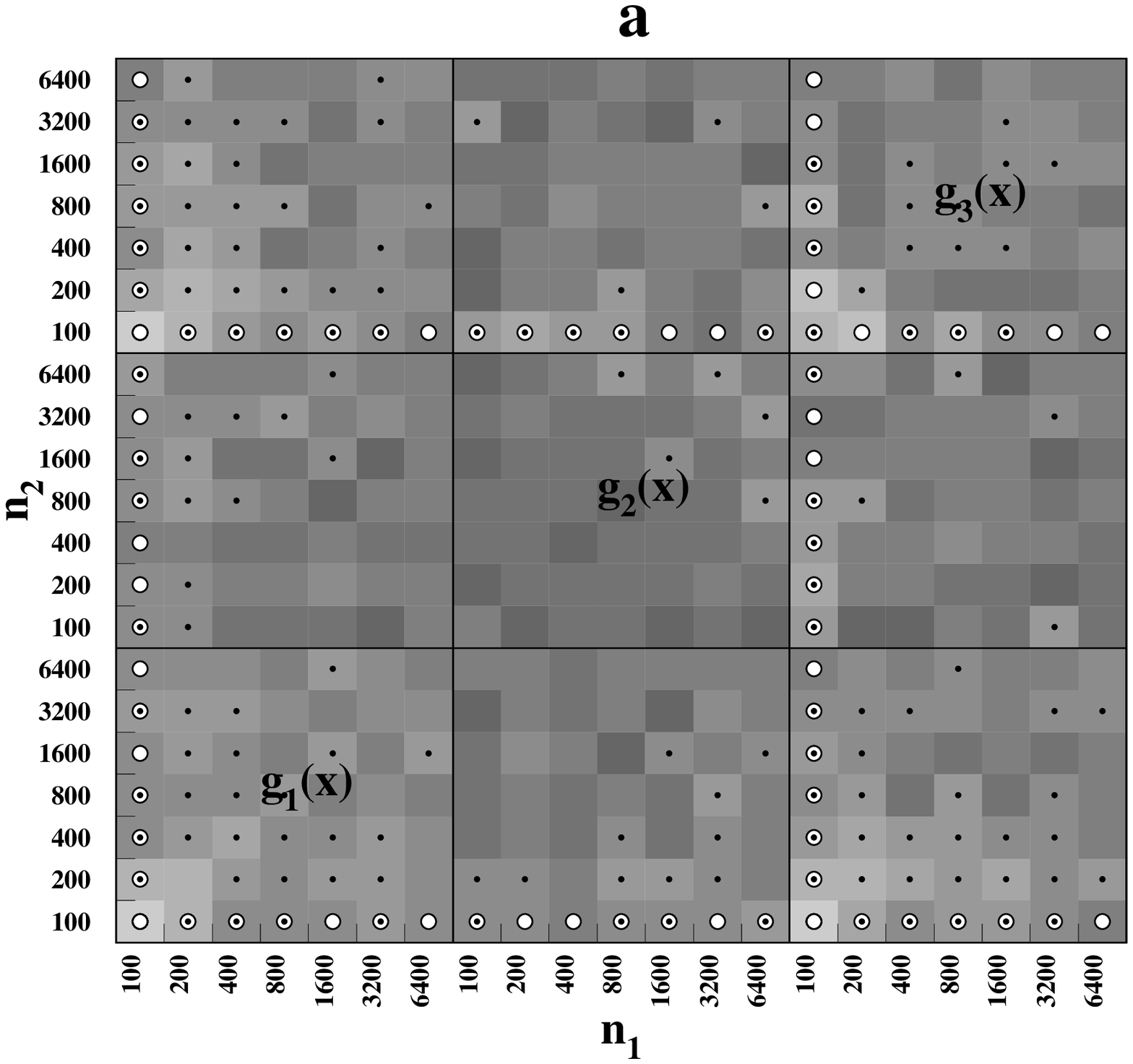}
\hspace*{-2.2 cm}  \includegraphics[width=4.5in]{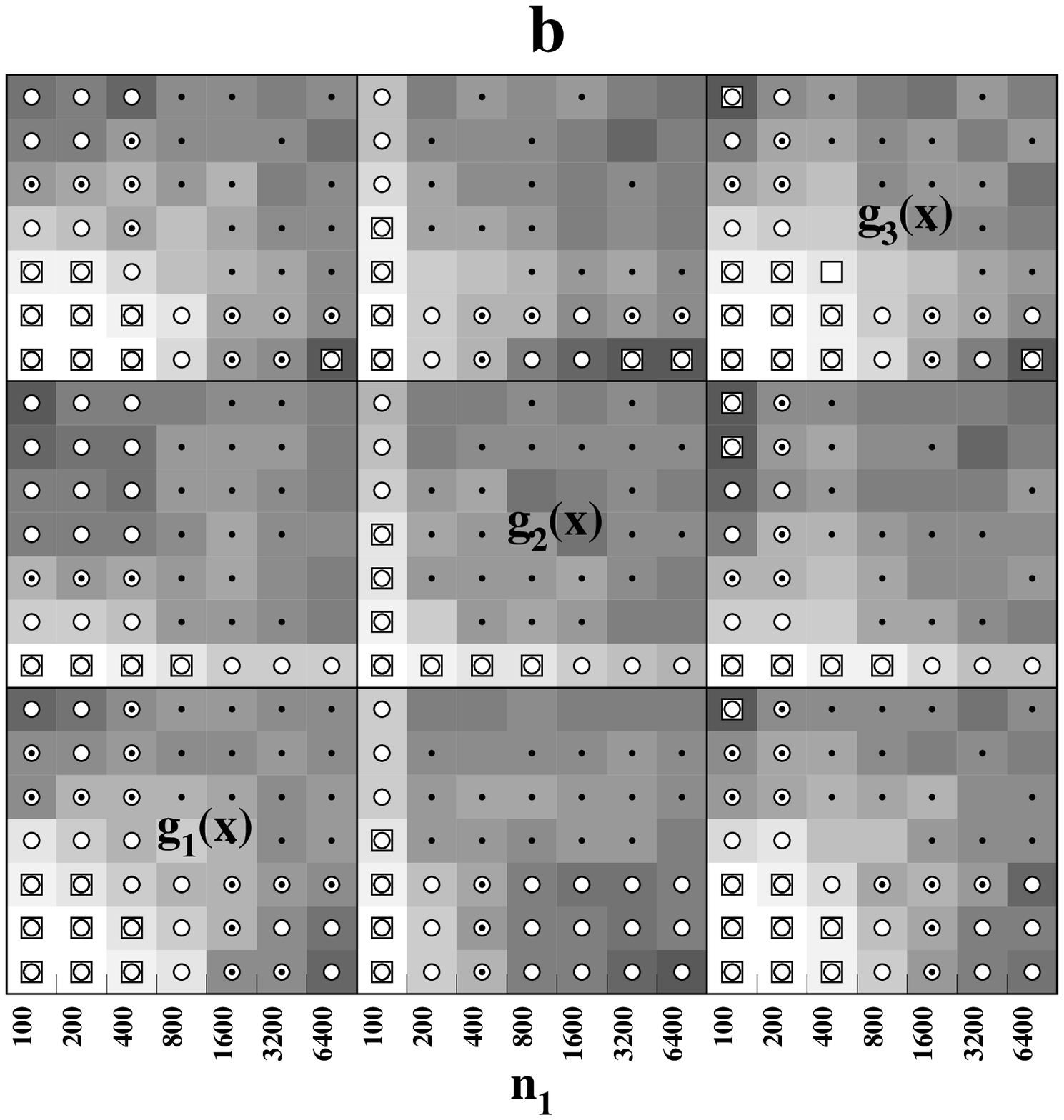} & \hspace*{-2.4 cm}\includegraphics[width=4.5in]{scalesize.eps}
\end{array}$
\end{center}
\vspace *{-1cm}
\hspace{1cm}\parbox{19cm}{ \caption{Sizes of the chi-square test for the comparison of two histograms with unnormalized weights: (a) number of bins $m=5$, (b) number of bins $m=20$. Markers show regions: ``$\circ$" have inappropriate number of events in the histograms for application of the test, $``\bullet"$ satisfy hypothesis $\alpha_s=0.05$, $``\Box"$ have a size of test $\alpha_s$ that is not close to the nominal size of the test.}}
\end{figure}
\end{landscape}

\begin{landscape}
\begin{figure}[h]
\begin{center}$
\begin{array}{ccc}
\includegraphics[width=4.5in]{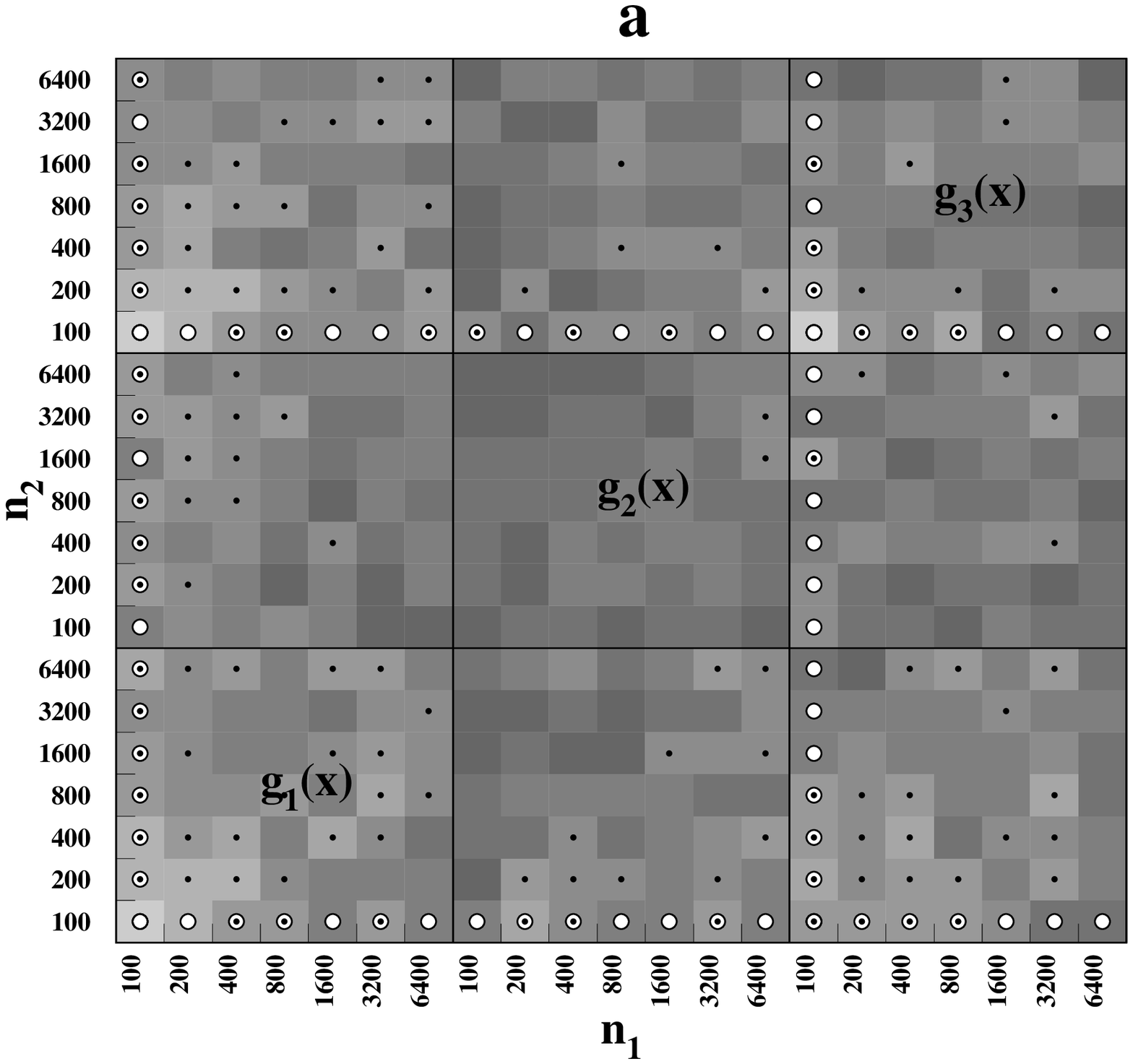}
\hspace*{-2.2 cm}  \includegraphics[width=4.5in]{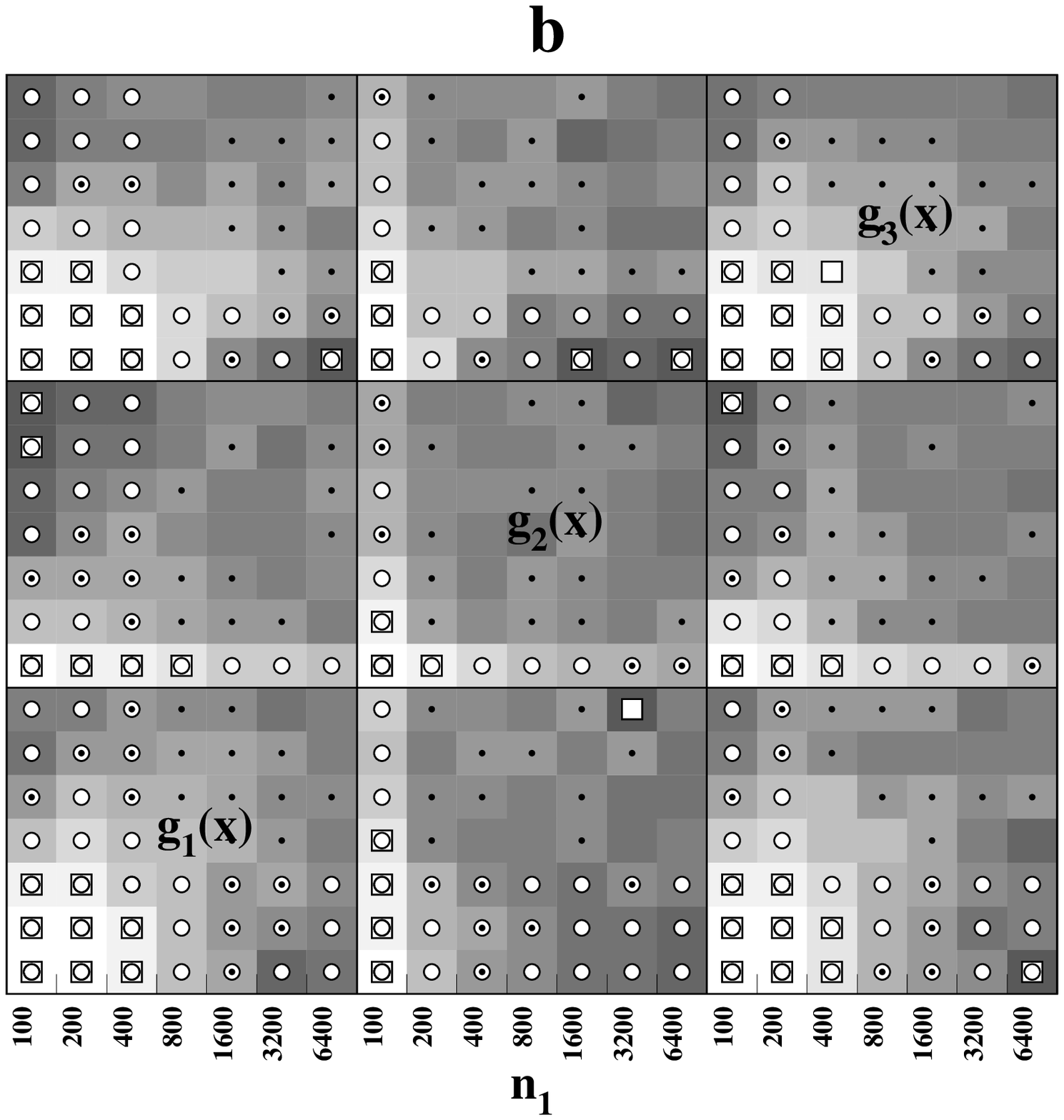} & \hspace*{-2.4 cm}\includegraphics[width=4.5in]{scalesize.eps}
\end{array}$
\end{center}
\vspace *{-1cm}
\hspace{1cm}\parbox{19cm}{\caption{Sizes of the chi-square test for the comparison histogram with normalized weights and histogram with unnormalized weights: (a) number of bins $m=5$, (b) number of bins $m=20$. Markers show regions: ``$\circ$" have inappropriate number of events in the histograms for application of the test, $``\bullet"$ satisfy hypothesis $\alpha_s=0.05$, $``\Box"$ have a size of test $alpha_s$ that is not close to the nominal size of the test.}}
\end{figure}
\end{landscape}

\clearpage
 \newpage
 \begin{landscape}
 \begin{table}\footnotesize
\parbox{19.5cm} {\caption{Sizes of the new test for the comparison of histogram with unweighted entries and histogram with unnormalised weights. Gray color of cell marks a size of test  with inappropriate number of events in the histograms; italic type marks a size of test satisfy hypothesis   $\alpha_s=0.05$; bold type marks a size of test that is not close to the nominal size of the test }}

\vspace *{0.3 cm}
\begin{minipage}{3.49in}
\begin{center}
\begin{tabular}{|c|>{$}l<{$}>{$}l<{$}>{$}l<{$}>{$}l<{$}>{$}l<{$}>{$}l<{$}>{$}l<{$}|  }
\hline
\multicolumn {1}{|c|}{} &  \multicolumn {7}{l|} {$\underline{m=5}  \,\,\,\,\,\qquad\qquad\qquad n_1$}\\
$n_2$ & 100 & 200 & 400 & 800 & 1600 & 3200 & 6400  \\ \hline
6400 & \cellcolor[gray]{.9} \textit{.054}   & {.056}  & {.055}  & {.056}  & {.056}  & \textit{.054}  & \textit{{.053}} \\
3200  & {\cellcolor[gray]{.9} .055}  & {.054}  & {.056}  & \textit{.054}  & \textit{.054}  & \textit{.052}  & \textit{.052} \\
1600  &\cellcolor[gray]{.9} \textit{.054}  & \textit{.053}  & \textit{.052}  & {.057}  & {.055}  & {.057}  & {.058} \\
800  &\cellcolor[gray]{.9} \textit{.051}  & \textit{.050}  & \textit{.050}  & \textit{.051}  & {.059}  & {.055}  & \textit{.053}  \\
400  &\cellcolor[gray]{.9} \textit{.052}  & \textit{.049}  & {.056}  & {.060}  & {.056}  & \textit{.051}  & {.058} \\
200  &\cellcolor[gray]{.9} \textit{.046}  & \textit{.046}  & \textit{.047}  & \textit{.052}  & \textit{.054}  & {.055}  & \textit{.052} \\
100  &\cellcolor[gray]{.9} .041  &\cellcolor[gray]{.9} {.046}  &\cellcolor[gray]{.9} \textit{.052}  &\cellcolor[gray]{.9} \textit{.053}  &\cellcolor[gray]{.9} {.054}  &\cellcolor[gray]{.9} {.054}  &\cellcolor[gray]{.9} \textit{.052} \\\hline
6400  &\cellcolor[gray]{.9} \textit{.052}  & {.055}  & \textit{.053}  & {.056}  & {.056}  & {.055}  & {.055} \\
3200  &\cellcolor[gray]{.9} \textit{.052}  & \textit{.052}  & \textit{.053}  & \textit{.052}  & {.058}  &{.059}  & {.055} \\
1600  &{\cellcolor[gray]{.9} .056}  & \textit{.051}  & \textit{.053}  &{.056}  & {.060}  & {.057}  &{.056} \\
800  &\cellcolor[gray]{.9} \textit{.051 } & \textit{.054}  & \textit{.054}  &{.057}  & {.060}  & {.055}  & {.058}  \\
400  &\cellcolor[gray]{.9} \textit{.053 } & {.055}  &{ .055}  & {.059}  & \textit{.053}  & {.058}  & {.056} \\
200  &\cellcolor[gray]{.9} \textit{.052}  & \textit{.053}  & {.056}  & {.061}  & {.057}  & {.060}  & {.056} \\
100  &{\cellcolor[gray]{.9} .056}  & {.055}  & {.057}  & {.055}  & {.056}  & {.062}  & {.060} \\\hline
6400  &\cellcolor[gray]{.9} \textit{.050}  & \textit{.053}  & \textit{.051}  & {.057}  & \textit{.052}  & \textit{.052}  & {.057} \\
3200  &\cellcolor[gray]{.9} \textit{.054}  & {.055}  & {.056}  & {.056}  & {.059}  & {.055}  & \textit{.054}  \\
1600  &\cellcolor[gray]{.9} \textit{.052}  & \textit{.053}  & {.056}  & {.057}  & \textit{.054}  & \textit{.051}  & {.054} \\
800  &\cellcolor[gray]{.9} \textit{.051}  & {.054}  & {.055}  & \textit{.052}  & {.055}  & \textit{.049}  & \textit{.053}  \\
400  &\cellcolor[gray]{.9} \textit{.046}  & \textit{.052}  & \textit{.049}  & {.055}  & \textit{.050}  & \textit{.053}  & {.058} \\
200  &\cellcolor[gray]{.9} \textit{.046}  & \textit{.047}  & \textit{.046}  & \textit{.053}  & {.057}  & {.055}  & {.056} \\
100  &\cellcolor[gray]{.9} .041  &\cellcolor[gray]{.9} .045  &\cellcolor[gray]{.9} \textit{.051}  &\cellcolor[gray]{.9} \textit{.051}  &{\cellcolor[gray]{.9} .055}  &\cellcolor[gray]{.9} \textit{.052}  &{\cellcolor[gray]{.9} .057} \\\hline
\end{tabular}
\end{center}
\end{minipage}
\begin{minipage}{3.49in}
\footnotesize
\begin{center}
\begin{tabular}{|>{$}l<{$}>{$}l<{$}>{$}l<{$}>{$}l<{$}>{$}l<{$}>{$}l<{$}>{$}l<{$}|>{$}c<{$}|  } \hline
\multicolumn {7}{|l|}{$\underline{m=20} \,\,\,\,\,\,\qquad\qquad\qquad n_1$} & \multicolumn {1}{c|}{}\\
100 & 200 & 400 & 800 & 1600 & 3200 & 6400 & $$w(x)$$ \\  \hline
\cellcolor[gray]{.9} {.061}  & \cellcolor[gray]{.9}{.056} & \cellcolor[gray]{.9}.055  & {.054}  & .056  & .056  & \textit{.054} &
\multirow{7}{*}{\large{$\frac{p(x)}{g_{3}(x)}$}}\\
\cellcolor[gray]{.9}{.061}  & \cellcolor[gray]{.9}.055  & \cellcolor[gray]{.9}.056  & .055  & \textit{.053}  & \textit{.054}  & \textit{.052} & \\
\cellcolor[gray]{.9}.056  & \cellcolor[gray]{.9}\textit{.050}  & \cellcolor[gray]{.9} \textit{.049}  & .055  & \textit{.050}  & \textit{.053}  & \textit{.049} & \\
\cellcolor[gray]{.9} .042  & \cellcolor[gray]{.9} .043  & \cellcolor[gray]{.9}.045  & .045  & \textit{.046}  & \textit{.052}  & .055 & \\
\cellcolor[gray]{.9} \textbf{.032} &\cellcolor[gray]{.9} \textbf{.033}  &\cellcolor[gray]{.9} .040  & .042  & .040  & \textit{.046}  & \textit{.052} & \\
\cellcolor[gray]{.9} \textbf{.025}  &\cellcolor[gray]{.9} \textbf{.026}  &\cellcolor[gray]{.9} \textbf{.028} &\cellcolor[gray]{.9} {.038}  &\cellcolor[gray]{.9} .044 &\cellcolor[gray]{.9} \textit{.047}  &\cellcolor[gray]{.9} \textit{.053} & \\
\cellcolor[gray]{.9} \textbf{.025}  & \cellcolor[gray]{.9} \textbf{.027 } &\cellcolor[gray]{.9} \textbf{.030}  &\cellcolor[gray]{.9} {.038}  &\cellcolor[gray]{.9} \textit{.053}  &\cellcolor[gray]{.9} .058  &\cellcolor[gray]{.9} \textbf{.065} & \\\hline
\cellcolor[gray]{.9} \textbf{.065}  &\cellcolor[gray]{.9} {.061}  &\cellcolor[gray]{.9} {.062}  & .055  & .055  & \textit{.054 } & .056 & \multirow{7}{*}{\large{$\frac{p(x)}{g_{2}(x)}$}}\\
\cellcolor[gray]{.9} \textbf{.064}  &\cellcolor[gray]{.9} .058  &\cellcolor[gray]{.9} .058  & .055  & \textit{.051}  & .058  & \textit{.053} & \\
\cellcolor[gray]{.9} {.062}  &\cellcolor[gray]{.9} .056  &\cellcolor[gray]{.9} .055  & \textit{.052}  & .056  & .056  & \textit{.051} & \\
%
\cellcolor[gray]{.9} {.062}  &\cellcolor[gray]{.9} \textit{.053}  &\cellcolor[gray]{.9} \textit{.050}  & .055  & .056  & .056  & \textit{.054} & \\

\cellcolor[gray]{.9} \textit{.048}  &\cellcolor[gray]{.9} \textit{.048}  &\cellcolor[gray]{.9} \textit{.050}  & \textit{.049}  & \textit{.053}  & .056  & .055 & \\
\cellcolor[gray]{.9} .043  &\cellcolor[gray]{.9} .044  &\cellcolor[gray]{.9} \textit{.047}  & \textit{.049}  & \textit{.050 } & \textit{.051}  & .056 & \\
\cellcolor[gray]{.9} \textbf{.025}  &\cellcolor[gray]{.9} \textbf{.033}  &\cellcolor[gray]{.9} \textbf{.034}  &\cellcolor[gray]{.9} \textbf{.036}  &\cellcolor[gray]{.9} .040  &\cellcolor[gray]{.9} .042  &\cellcolor[gray]{.9} .043 & \\\hline
\cellcolor[gray]{.9} .056  &\cellcolor[gray]{.9} .057  &\cellcolor[gray]{.9} \textit{.052}  & \textit{.054}  & \textit{.053}  & .058  & .055 &\multirow{7}{*}{\large{$\frac{p(x)}{g_{1}(x)}$} \normalsize{$=1$}} \\
\cellcolor[gray]{.9} .059  &\cellcolor[gray]{.9} \textit{.051}  &\cellcolor[gray]{.9} \textit{.051}  & \textit{.049}  & \textit{.049}  & \textit{.052}  & .056 & \\
\cellcolor[gray]{.9} \textit{.051}  &\cellcolor[gray]{.9} .044  &\cellcolor[gray]{.9} \textit{.047}  & \textit{.047 } & \textit{.049}  & \textit{.054}  & \textit{.054} & \\
%
%
\cellcolor[gray]{.9} .044  &\cellcolor[gray]{.9} .040  &\cellcolor[gray]{.9} .043  & .044  & \textit{.047 } & \textit{.052}  & .057 &  \\
\cellcolor[gray]{.9} \textbf{.031}  &\cellcolor[gray]{.9} \textbf{.034}  &\cellcolor[gray]{.9} .041  &\cellcolor[gray]{.9} .043  &\cellcolor[gray]{.9} \textit{.050}  &\cellcolor[gray]{.9} \textit{.050}  &\cellcolor[gray]{.9} .055 & \\
\cellcolor[gray]{.9} \textbf{.025}  &\cellcolor[gray]{.9} \textbf{.027}  &\cellcolor[gray]{.9} \textbf{.034}  &\cellcolor[gray]{.9} .044  &\cellcolor[gray]{.9} \textit{.051}  &\cellcolor[gray]{.9} \textit{.054}  &\cellcolor[gray]{.9} .056 &\\
\cellcolor[gray]{.9} \textbf{.025}  &\cellcolor[gray]{.9} \textbf{.026}  &\cellcolor[gray]{.9} \textbf{.032}  &\cellcolor[gray]{.9} .045  &\cellcolor[gray]{.9} \textit{.048 } &\cellcolor[gray]{.9} {.061}  &\cellcolor[gray]{.9} .060 & \\\hline
\end{tabular}
\end{center}
\end{minipage}
\end{table}
\end{landscape}
\clearpage
 \newpage

\begin{landscape}
 \begin{table}\footnotesize
\parbox{20.5cm} {\caption{Sizes of the heuristic test  for the comparison of  histogram with unweighted entries and histogram with unnormalized weights.  Gray color of cell marks a size of test  with inappropriate number of events in the histograms; italic type marks a size of test satisfy hypothesis $\alpha_s=0.05$; bold type marks a size of test that is not close to the nominal size of the test}}

\vspace *{0.3 cm}
\begin{minipage}{3.74in}
\begin{center}
\begin{tabular}{|c|>{$}l<{$}>{$}l<{$}>{$}l<{$}>{$}l<{$}>{$}l<{$}>{$}l<{$}>{$}l<{$}|  }
\hline
\multicolumn {1}{|c|}{} &  \multicolumn {7}{l|} {$\underline{m=5}\,\, \,\,\,\,\, \,\qquad\qquad\qquad n_1$}\\
$n_2$ & 100 & 200 & 400 & 800 & 1600 & 3200 & 6400  \\ \hline
6400  & \cellcolor[gray]{.9} \textit{.054}  &\cellcolor[gray]{.9} .055  &\cellcolor[gray]{.9} \textit{.050}  & \textit{.054}  & \textit{.054}  & \textit{.054}  & .055 \\
3200  & \cellcolor[gray]{.9} \textit{.051}  & \cellcolor[gray]{.9} \textit{.053}  & \cellcolor[gray]{.9} .058  & \textit{.049}  & .057  & \textit{.053 } & .061  \\
 1600 &\cellcolor[gray]{.9} \textit{.050}  &\cellcolor[gray]{.9} \textit{.047}  &\cellcolor[gray]{.9} \textit{.050}  & .058  & .058  & {.061}  & .060 \\
800 &\cellcolor[gray]{.9} \textit{.053}  &\cellcolor[gray]{.9} \textit{.054}  &\cellcolor[gray]{.9} \textit{.053}  & .059  & {.062}  & {.061}  & {.062 } \\
400  &\cellcolor[gray]{.9} \textbf{.064}  &\cellcolor[gray]{.9} .060  &\cellcolor[gray]{.9} \textbf{.064}  &\cellcolor[gray]{.9} \textbf{.066}  &\cellcolor[gray]{.9} \textbf{.065}  &\cellcolor[gray]{.9} {.061}  &\cellcolor[gray]{.9} \textbf{.064}  \\
200  &\cellcolor[gray]{.9} \textbf{.071}  &\cellcolor[gray]{.9} \textbf{.077}  &\cellcolor[gray]{.9} \textbf{.071}  &\cellcolor[gray]{.9} {.062}  &\cellcolor[gray]{.9} {.063}  &\cellcolor[gray]{.9} {.063}  &\cellcolor[gray]{.9} \textbf{.069}  \\
100  &\cellcolor[gray]{.9} \textbf{.093}  &\cellcolor[gray]{.9} \textbf{.086}  &\cellcolor[gray]{.9} \textbf{.077}  &\cellcolor[gray]{.9} \textbf{.072}  &\cellcolor[gray]{.9} \textbf{.068}  &\cellcolor[gray]{.9} {.062}  &\cellcolor[gray]{.9} \textbf{.068}  \\\hline
6400  &\cellcolor[gray]{.9} \textit{.048}  &\cellcolor[gray]{.9} \textit{.051}  &\cellcolor[gray]{.9} \textit{.051}  & \textit{.049}  & \textit{.048}  & \textit{.052}  & \textit{.054}  \\
3200  &\cellcolor[gray]{.9} \textit{.050}  &\cellcolor[gray]{.9} \textit{.052}  &\cellcolor[gray]{.9} .050  & .046  & .050  & .055  & .057  \\
 1600 &\cellcolor[gray]{.9} \textit{.052}  &\cellcolor[gray]{.9} \textit{.051}  &\cellcolor[gray]{.9} .055  & .058  & \textit{.054}  & \textbf{.064}  & {.062}  \\
800  &\cellcolor[gray]{.9} \textit{.050}  &\cellcolor[gray]{.9} \textit{.051}  &\cellcolor[gray]{.9} .055  & .057  & \textbf{.067}  & {.061}  & \textbf{.070} \\
400  &\cellcolor[gray]{.9} \textit{.054}  &\cellcolor[gray]{.9} .060  &\cellcolor[gray]{.9} {.063}  & \textbf{.073}  & \textbf{.064}  & \textbf{.068}  & \textbf{.070}  \\
200  &\cellcolor[gray]{.9} \textbf{.064}  &\cellcolor[gray]{.9} \textbf{.068}  &\cellcolor[gray]{.9} \textbf{.078}  & \textbf{.071}  & \textbf{.072 } & \textbf{.074}  & \textbf{.075}  \\
100  &\cellcolor[gray]{.9} \textbf{.078}  &\cellcolor[gray]{.9} \textbf{.097}  &\cellcolor[gray]{.9} \textbf{.082}  &\cellcolor[gray]{.9} \textbf{.079 } &\cellcolor[gray]{.9} \textbf{.077}  &\cellcolor[gray]{.9} \textbf{.080}  &\cellcolor[gray]{.9} \textbf{.079}  \\\hline
6400  &\cellcolor[gray]{.9} \textit{.054}  &\cellcolor[gray]{.9} \textit{.052}  &\cellcolor[gray]{.9} \textit{.052}  & \textit{.046}  & .056  & \textit{.048}  & \textit{.052}  \\
3200  &\cellcolor[gray]{.9} .055  &\cellcolor[gray]{.9} \textit{.051}  &\cellcolor[gray]{.9} \textit{.051}  & \textit{.048}  & \textit{.049}  & \textit{.054}  & \textit{.054}  \\
1600  &\cellcolor[gray]{.9} \textit{.054}  &\cellcolor[gray]{.9} \textit{.054 } &\cellcolor[gray]{.9} \textit{.051}  & \textit{.054}  & \textit{.053}  & \textit{.048}  & \textit{.049 } \\
800  &\cellcolor[gray]{.9} .056  &\cellcolor[gray]{.9} .058  &\cellcolor[gray]{.9} .060  & .058  & .056  & .055  & \textit{.049}  \\
400  &\cellcolor[gray]{.9} \textbf{.066}  &\cellcolor[gray]{.9} \textbf{.065}  &\cellcolor[gray]{.9} \textbf{.070}  &\cellcolor[gray]{.9} .055  &\cellcolor[gray]{.9} \textit{.054}  &\cellcolor[gray]{.9} \textit{.053}  &\cellcolor[gray]{.9} \textit{.051}  \\
200  &\cellcolor[gray]{.9} \textbf{.075}  &\cellcolor[gray]{.9} \textbf{.083}  &\cellcolor[gray]{.9} \textbf{.064}  &\cellcolor[gray]{.9} .055  &\cellcolor[gray]{.9} \textit{.053}  &\cellcolor[gray]{.9} \textit{.053 } &\cellcolor[gray]{.9} \textit{.048}  \\
 100 &\cellcolor[gray]{.9} \textbf{.083}  &\cellcolor[gray]{.9} \textbf{.078}  &\cellcolor[gray]{.9} {.062}  &\cellcolor[gray]{.9} .055  &\cellcolor[gray]{.9} \textit{.051}  &\cellcolor[gray]{.9} \textit{.047}  &\cellcolor[gray]{.9} \textit{.050 } \\\hline
\end{tabular}
\end{center}
\end{minipage}
\vspace{-0.2cm}
\begin{minipage}{3.74in}
\begin{center}
\begin{tabular}{|>{$}l<{$}>{$}l<{$}>{$}l<{$}>{$}l<{$}>{$}l<{$}>{$}l<{$}>{$}l<{$}|>{$}c<{$}|  } \hline
\multicolumn {7}{|l|}{$\underline{m=20} \, \,\,\,\,\,\,\,\qquad\qquad\qquad n_1$} & \multicolumn {1}{c|}{}\\
100 & 200 & 400 & 800 & 1600 & 3200 & 6400 & $$w(x)$$ \\  \hline
\cellcolor[gray]{.9} {.063}  &\cellcolor[gray]{.9} .060  &\cellcolor[gray]{.9} .059  &\cellcolor[gray]{.9} \textit{.053}  &\cellcolor[gray]{.9} .059  & .058  & {.061} & \multirow{7}{*}{\large{$\frac{p(x)}{g_{3}(x)}$}}\\
 \cellcolor[gray]{.9} \textbf{.069}  &\cellcolor[gray]{.9} {.062}  &\cellcolor[gray]{.9} {.061}  &\cellcolor[gray]{.9} {.062}  &\cellcolor[gray]{.9} {.062}  & \textbf{.067}  & {.062} & \\
 \cellcolor[gray]{.9} \textbf{.084}  &\cellcolor[gray]{.9} \textbf{.076}  &\cellcolor[gray]{.9} \textbf{.078}  &\cellcolor[gray]{.9} \textbf{.085 } &\cellcolor[gray]{.9} \textbf{.081}  & \textbf{.071}  & \textbf{.067} & \\
\cellcolor[gray]{.9} \textbf{.112}  &\cellcolor[gray]{.9} \textbf{.109}  &\cellcolor[gray]{.9} \textbf{.114}  &\cellcolor[gray]{.9} \textbf{.121}  &\cellcolor[gray]{.9} \textbf{.089}  &\cellcolor[gray]{.9} \textbf{.069 } &\cellcolor[gray]{.9} \textbf{.068} & \\
 \cellcolor[gray]{.9} \textbf{.136}  &\cellcolor[gray]{.9} \textbf{.147}  &\cellcolor[gray]{.9} \textbf{.151}  &\cellcolor[gray]{.9} \textbf{.126 } &\cellcolor[gray]{.9} \textbf{.087 } &\cellcolor[gray]{.9} \textbf{.066}  &\cellcolor[gray]{.9} \textbf{.065} & \\
\cellcolor[gray]{.9} \textbf{.158}  &\cellcolor[gray]{.9} \textbf{.155}  &\cellcolor[gray]{.9} \textbf{.148}  &\cellcolor[gray]{.9} \textbf{.123 } &\cellcolor[gray]{.9} \textbf{.083}  &\cellcolor[gray]{.9} \textbf{.067}  &\cellcolor[gray]{.9} .060 & \\
\cellcolor[gray]{.9} \textbf{.137}  &\cellcolor[gray]{.9} \textbf{.122}  &\cellcolor[gray]{.9} \textbf{.175}  &\cellcolor[gray]{.9} \textbf{.151}  &\cellcolor[gray]{.9} \textbf{.104}  &\cellcolor[gray]{.9} \textbf{.084 } &\cellcolor[gray]{.9} \textbf{.076 }& \\\hline
 \cellcolor[gray]{.9} .060  &\cellcolor[gray]{.9} .057  &\cellcolor[gray]{.9} .058  &\cellcolor[gray]{.9} .056  &\cellcolor[gray]{.9} \textit{.052}  & .055  & \textit{.054} & \multirow{7}{*}{\large{$\frac{p(x)}{g_{2}(x)}$}}\\
  \cellcolor[gray]{.9} {.063 } &\cellcolor[gray]{.9} \textbf{.065 } &\cellcolor[gray]{.9} .060  &\cellcolor[gray]{.9} \textit{.050}  &\cellcolor[gray]{.9} .057  & .054  & {.062} & \\
  \cellcolor[gray]{.9} \textbf{.067}  &\cellcolor[gray]{.9} {.062}  &\cellcolor[gray]{.9} \textbf{.064}  &\cellcolor[gray]{.9} {.062}  &\cellcolor[gray]{.9} \textbf{.065}  & \textbf{.078}  & \textbf{.068} & \\
  \cellcolor[gray]{.9} \textbf{.074}  &\cellcolor[gray]{.9} \textbf{.070}  &\cellcolor[gray]{.9} \textbf{.078}  &\cellcolor[gray]{.9} \textbf{.077}  &\cellcolor[gray]{.9} \textbf{.097}  & \textbf{.075}  & \textbf{.071} &\\
  \cellcolor[gray]{.9} \textbf{.104}  &\cellcolor[gray]{.9} \textbf{.101}  &\cellcolor[gray]{.9} \textbf{.108 } &\cellcolor[gray]{.9} \textbf{.140}  &\cellcolor[gray]{.9} \textbf{.105}  &\cellcolor[gray]{.9} \textbf{.084}  &\cellcolor[gray]{.9} \textbf{.073} & \\
  \cellcolor[gray]{.9} \textbf{.162}  &\cellcolor[gray]{.9} \textbf{.184}  &\cellcolor[gray]{.9} \textbf{.140}  &\cellcolor[gray]{.9} \textbf{.135}  &\cellcolor[gray]{.9} \textbf{.098}  &\cellcolor[gray]{.9} \textbf{.081}  &\cellcolor[gray]{.9} \textbf{.076} & \\
  \cellcolor[gray]{.9} \textbf{.290}  &\cellcolor[gray]{.9} \textbf{.115 } &\cellcolor[gray]{.9} \textbf{.134}  &\cellcolor[gray]{.9} \textbf{.130 } &\cellcolor[gray]{.9} \textbf{.090}  &\cellcolor[gray]{.9} \textbf{.070 } &\cellcolor[gray]{.9} \textbf{.064} & \\\hline
  \cellcolor[gray]{.9} \textbf{.064}  &\cellcolor[gray]{.9} {.061}  &\cellcolor[gray]{.9} .058  &\cellcolor[gray]{.9} .055  &\cellcolor[gray]{.9} .058  & \textit{.052}  & .057 &\multirow{7}{*}{\large{$\frac{p(x)}{g_{1}(x)}$} \normalsize{$=1$}} \\
  \cellcolor[gray]{.9} \textbf{.068}  &\cellcolor[gray]{.9} .059  &\cellcolor[gray]{.9} \textbf{.070}  &\cellcolor[gray]{.9} \textbf{.069}  &\cellcolor[gray]{.9} \textbf{.068}  & \textbf{.069}  & .060 & \\
  \cellcolor[gray]{.9} \textbf{.086}  &\cellcolor[gray]{.9} \textbf{.081}  &\cellcolor[gray]{.9} \textbf{.082}  &\cellcolor[gray]{.9} \textbf{.083 } &\cellcolor[gray]{.9} \textbf{.092 } &\cellcolor[gray]{.9} \textbf{.065}  &\cellcolor[gray]{.9} .058 & \\
  \cellcolor[gray]{.9} \textbf{.113}  &\cellcolor[gray]{.9} \textbf{.110}  &\cellcolor[gray]{.9} \textbf{.121}  &\cellcolor[gray]{.9} \textbf{.123}  &\cellcolor[gray]{.9} \textbf{.087}  &\cellcolor[gray]{.9} \textbf{.065 } &\cellcolor[gray]{.9} .055 & \\
  \cellcolor[gray]{.9} \textbf{.139}  &\cellcolor[gray]{.9} \textbf{.128 } &\cellcolor[gray]{.9} \textbf{.136}  &\cellcolor[gray]{.9} \textbf{.115}  &\cellcolor[gray]{.9} \textbf{.080}  &\cellcolor[gray]{.9} .058  &\cellcolor[gray]{.9} \textit{.054} & \\
  \cellcolor[gray]{.9} \textbf{.137}  &\cellcolor[gray]{.9} \textbf{.141}  &\cellcolor[gray]{.9} \textbf{.136}  &\cellcolor[gray]{.9} \textbf{.119 } &\cellcolor[gray]{.9} \textbf{.084}  &\cellcolor[gray]{.9} {.063}  &\cellcolor[gray]{.9} .056 & \\
  \cellcolor[gray]{.9} \textbf{.128}  &\cellcolor[gray]{.9} \textbf{.128}  &\cellcolor[gray]{.9} \textbf{.169}  &\cellcolor[gray]{.9} \textbf{.152}  &\cellcolor[gray]{.9} \textbf{.112}  &\cellcolor[gray]{.9} \textbf{.089}  &\cellcolor[gray]{.9} \textbf{.073} & \\\hline
\end{tabular}
\end{center}
\end{minipage}
\end{table}
\end{landscape}

\newpage
The powers of the tests were investigated for slightly different values of the amplitude of the second  peak of the specified probability  distribution function (see Fig. 5):
\begin{equation}
p_0(x)\propto \frac{2}{(x-10)^2+1}+\frac{1.15}{(x-14)^2+1}. \label{weightr}
\end{equation}

The results of these calculations are presented in Figs. 6-8. All the plots were designated the same scale to facilitate comparison. It must be noted that the powers of the tests for histograms with unnormalized wights(see Figs. 7-8) are lower than those of the test for the histogram with normalized weights. It can be observed  that the powers of the tests for histograms with 5 bins are greater than those with 20 bins, except for the comparison of the two histograms with normalized weights and the function $g_3(x)$. It can be explained that for the case of 20 bins in region where the histograms differ we have more detailed information about the shape but it is represented by bins with small statistics of events. In addition, the power is large for the pairs of histograms that have the function $g_3(x)$ in one of them because more events are generated in region where the histograms differ; this is in agreement with the results presented in Ref. \cite {gagunashvili3}. The comparison of the powers of the tests for different pairs of the histograms  with the powers of the tests for the histograms with unweighted entries   demonstrated that the values of the powers as well as the sizes of the new tests are  reasonable.

\begin{figure}
\centering \vspace*{-3cm}
\hspace*{-0.7 cm}
  \includegraphics[width=1.15 \textwidth]{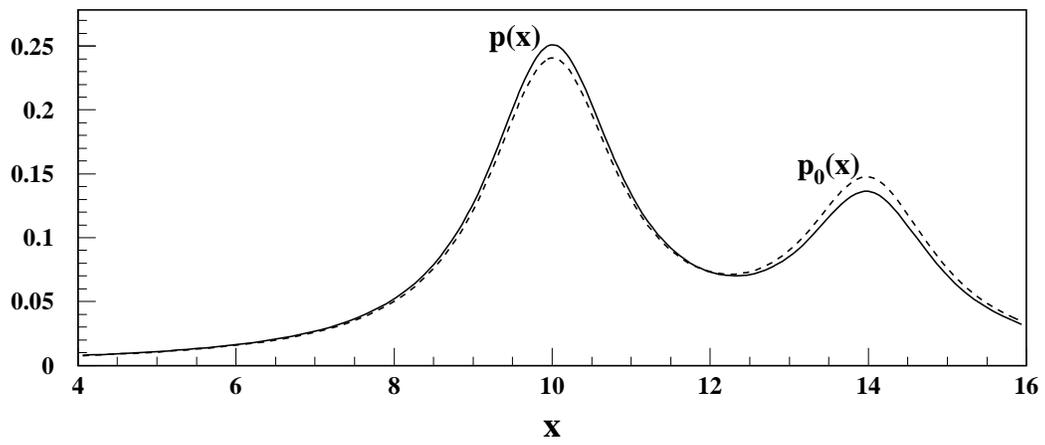}\\
\vspace*{-8. cm}
\caption {Probability density function $p(x)$
(solid line)  and $p_0(x)$ (dashed line).} \vspace*{3 cm }
\end{figure}

\begin{landscape}
\begin{figure}[h]
\begin{center}$
\begin{array}{ccc}
\includegraphics[width=4.5in]{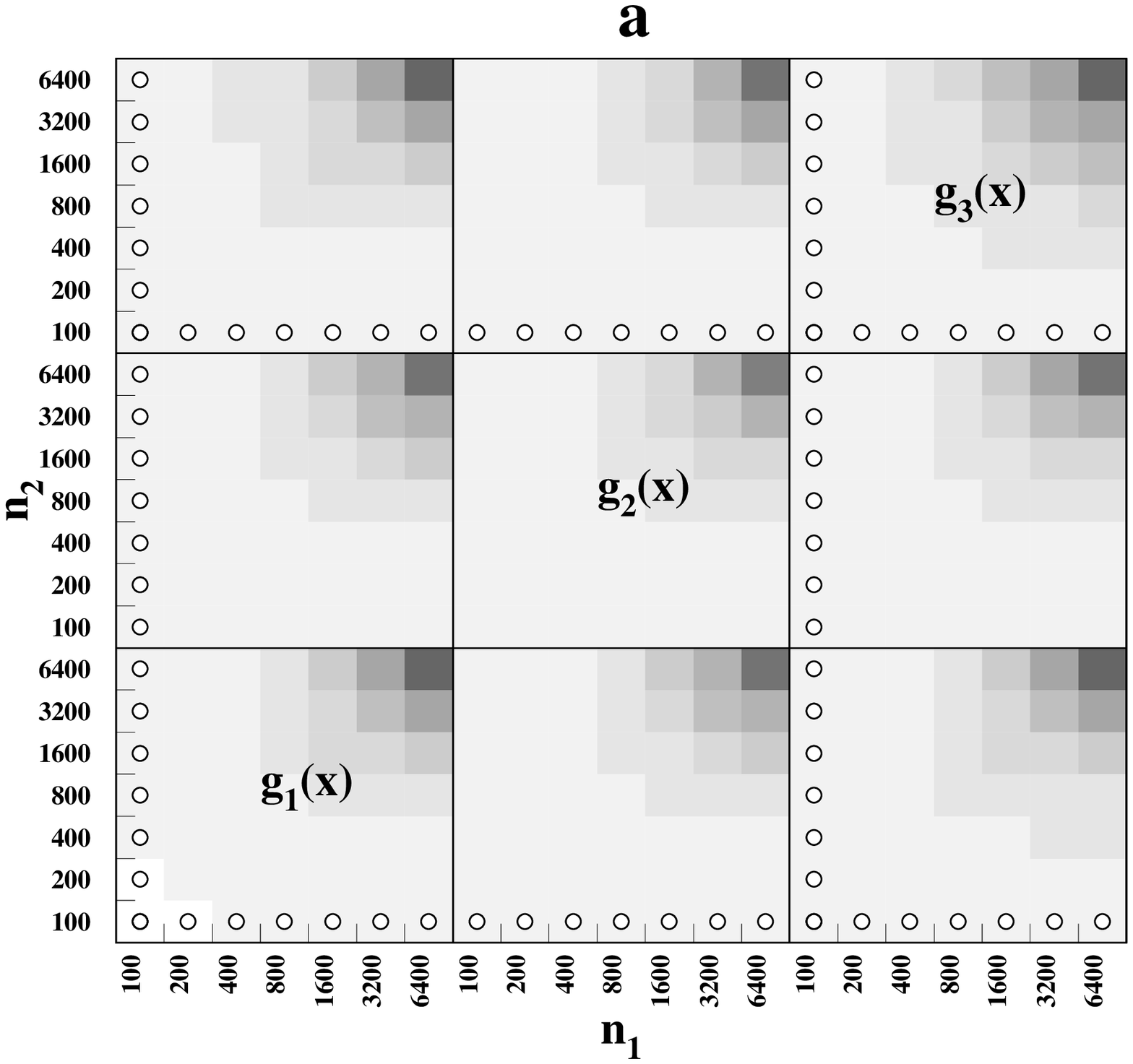}
\hspace*{-2.2 cm}  \includegraphics[width=4.5in]{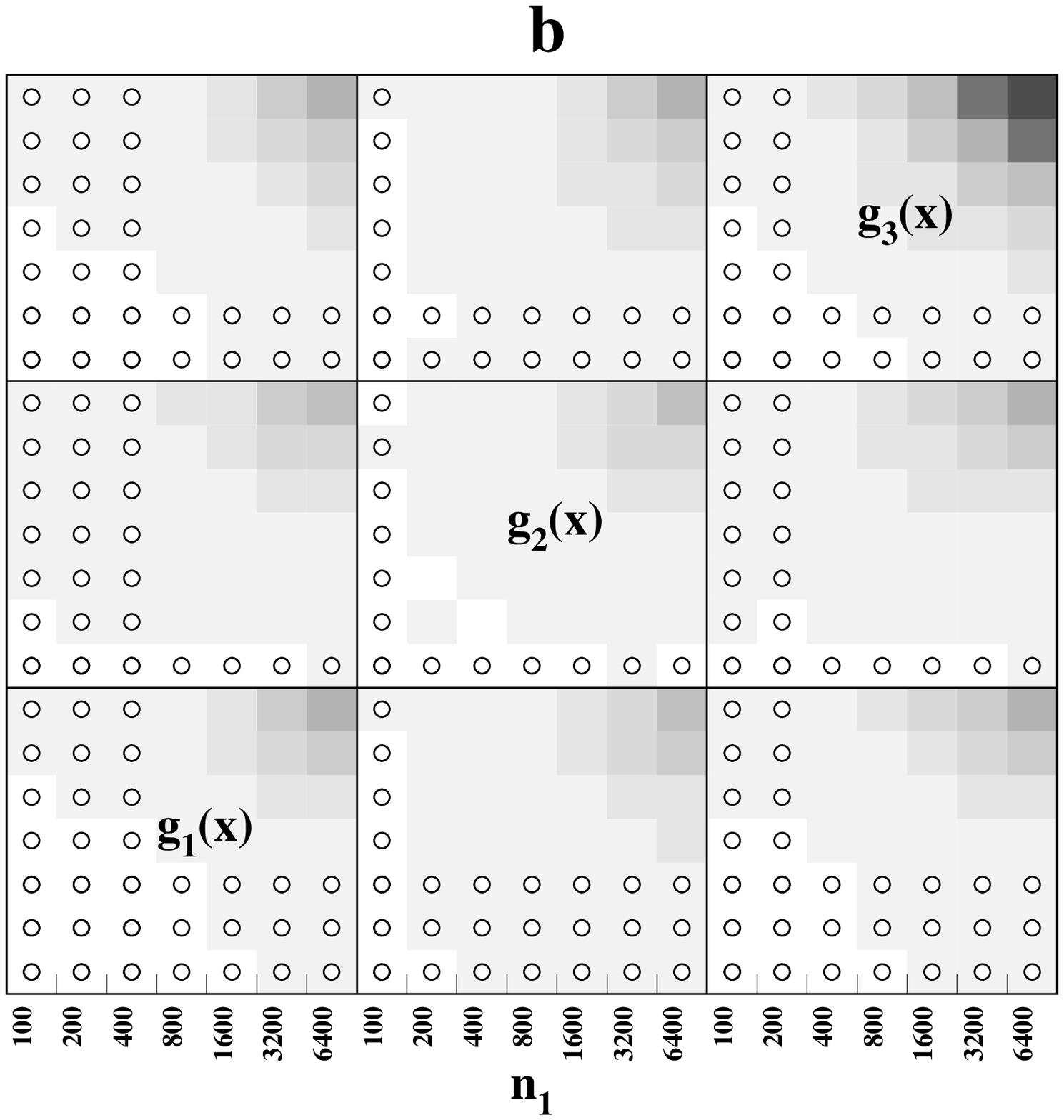} & \hspace*{-2.4 cm}\includegraphics[width=4.5in]{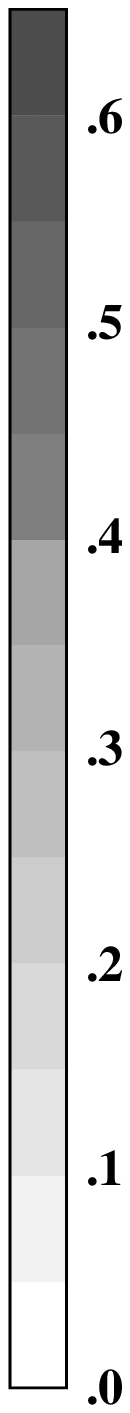}
\end{array}$
\end{center}
\vspace *{-1cm}
\hspace{1cm} \parbox{19cm}{\caption{Powers of the chi-square test for the comparison of two weighted histograms with normalized weights: (a) number of bins $m=5$, (b) number of bins $m=20$. Markers show regions with inappropriate number of events in the histograms for application of the test.}}
\end{figure}
\end{landscape}

\begin{landscape}
\begin{figure}[h]
\begin{center}$
\begin{array}{ccc}
\includegraphics[width=4.5in]{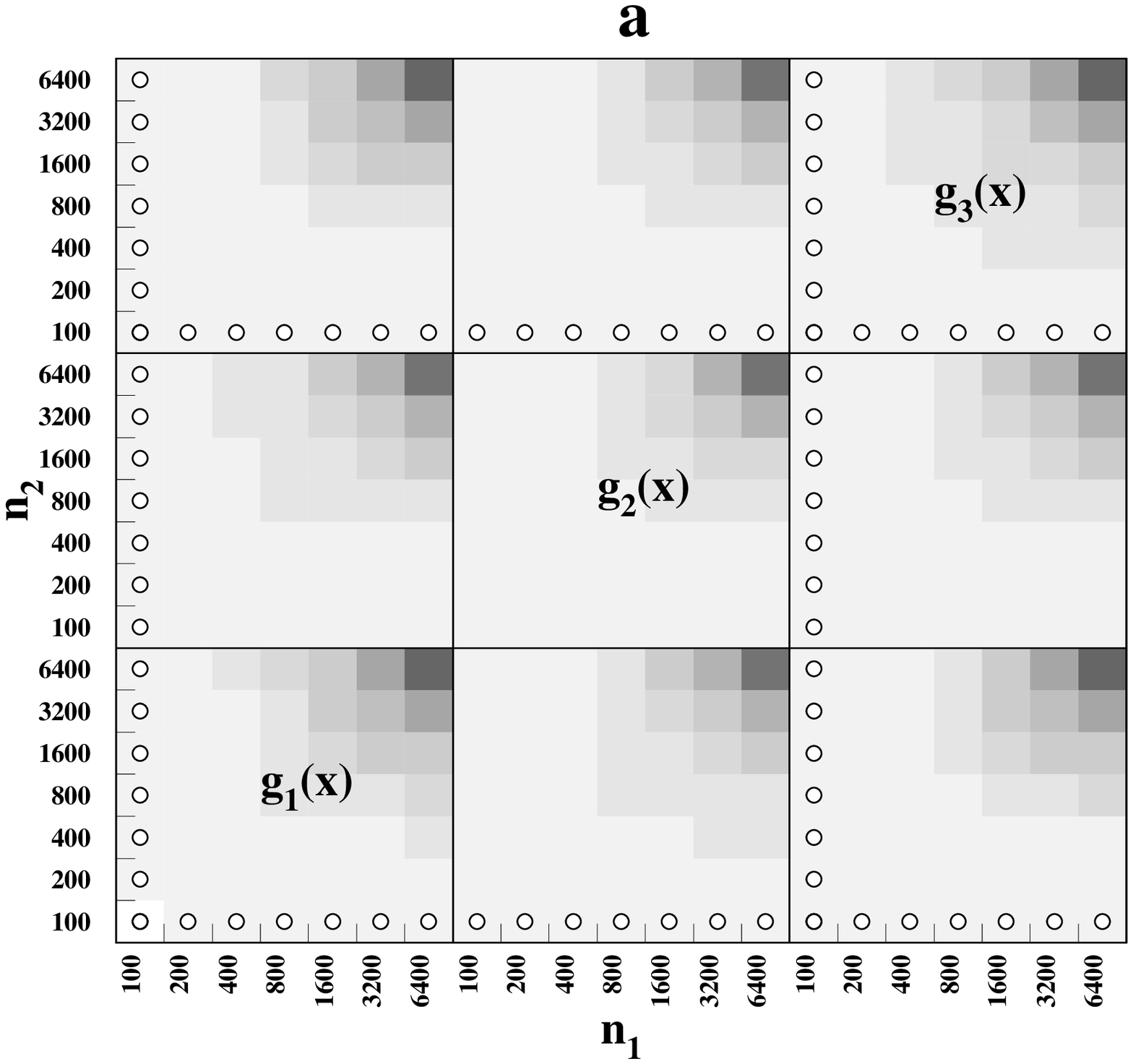}
\hspace*{-2.2 cm}  \includegraphics[width=4.5in]{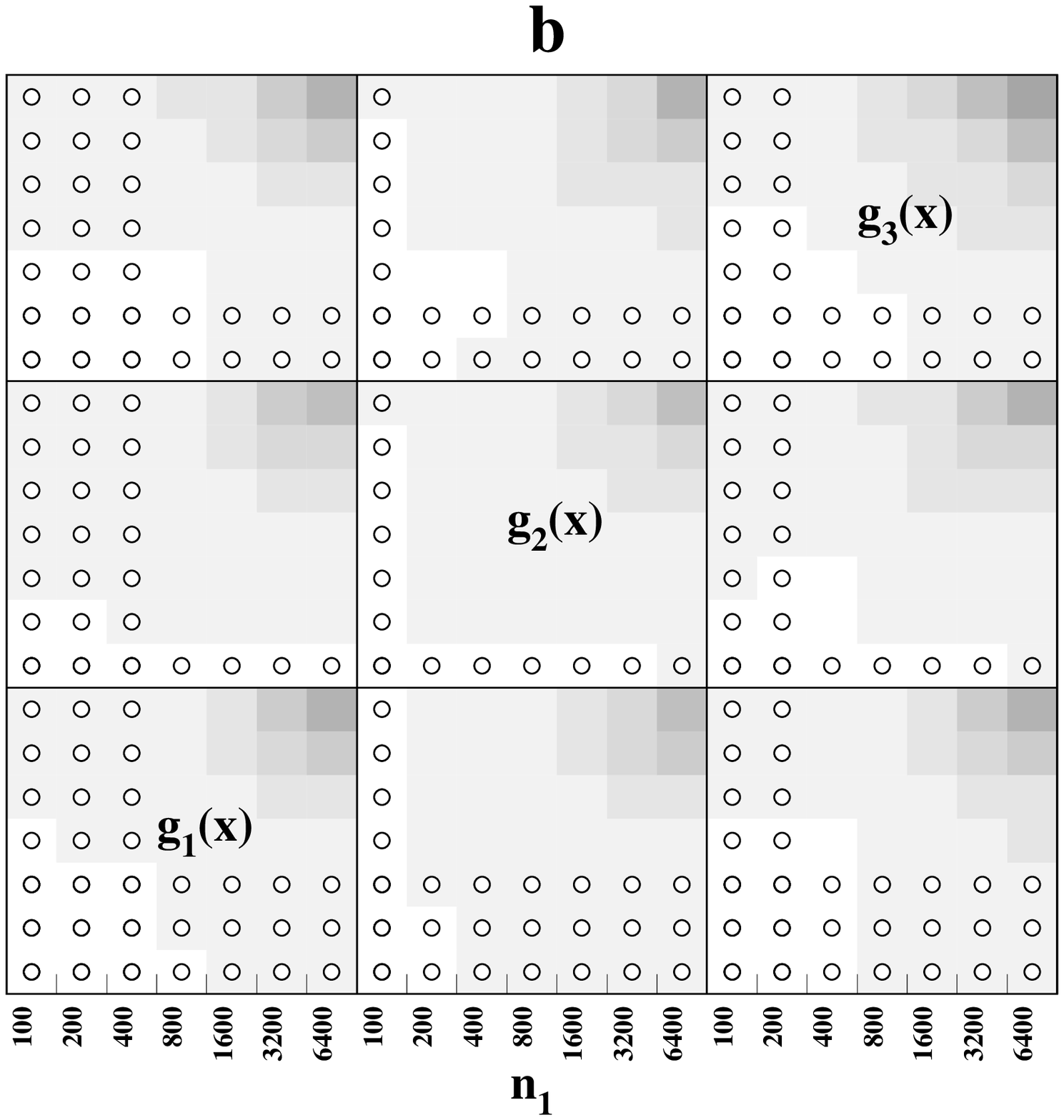} & \hspace*{-2.4 cm}\includegraphics[width=4.5in]{scalesizep.eps}
\end{array}$
\end{center}
\vspace *{-1cm}
\hspace{1cm}\parbox{19cm}{\caption{ Powers of the chi-square test for the comparison of two histograms  with unnormalized weights: (a) number of bins $m=5$, (b) number of bins $m=20$. Markers show regions with inappropriate number of events in the histograms for application of the test.}}
\end{figure}
\end{landscape}

\begin{landscape}
\begin{figure}[h]
\begin{center}$
\begin{array}{ccc}
\includegraphics[width=4.5in]{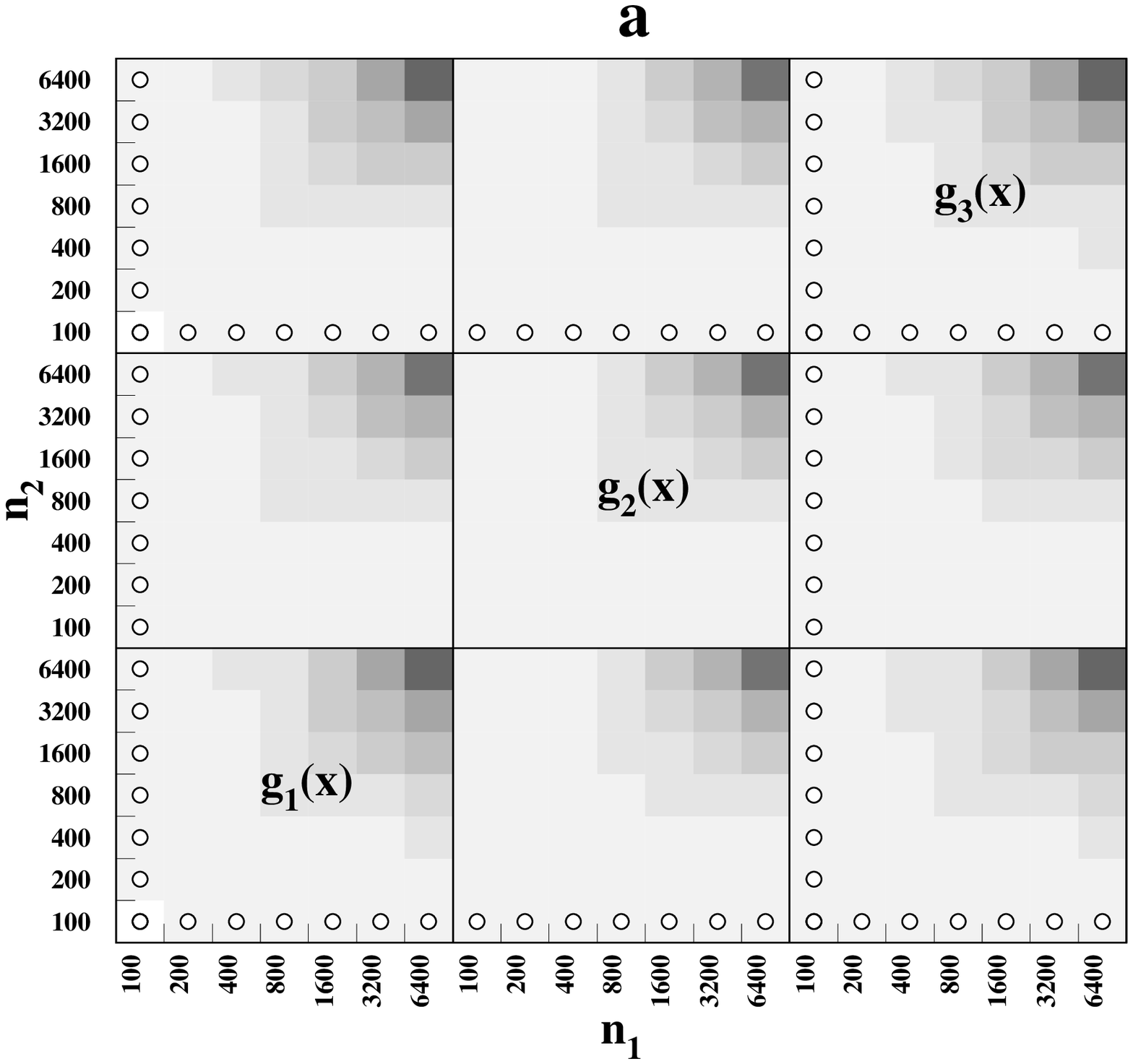}
\hspace*{-2.2 cm}  \includegraphics[width=4.5in]{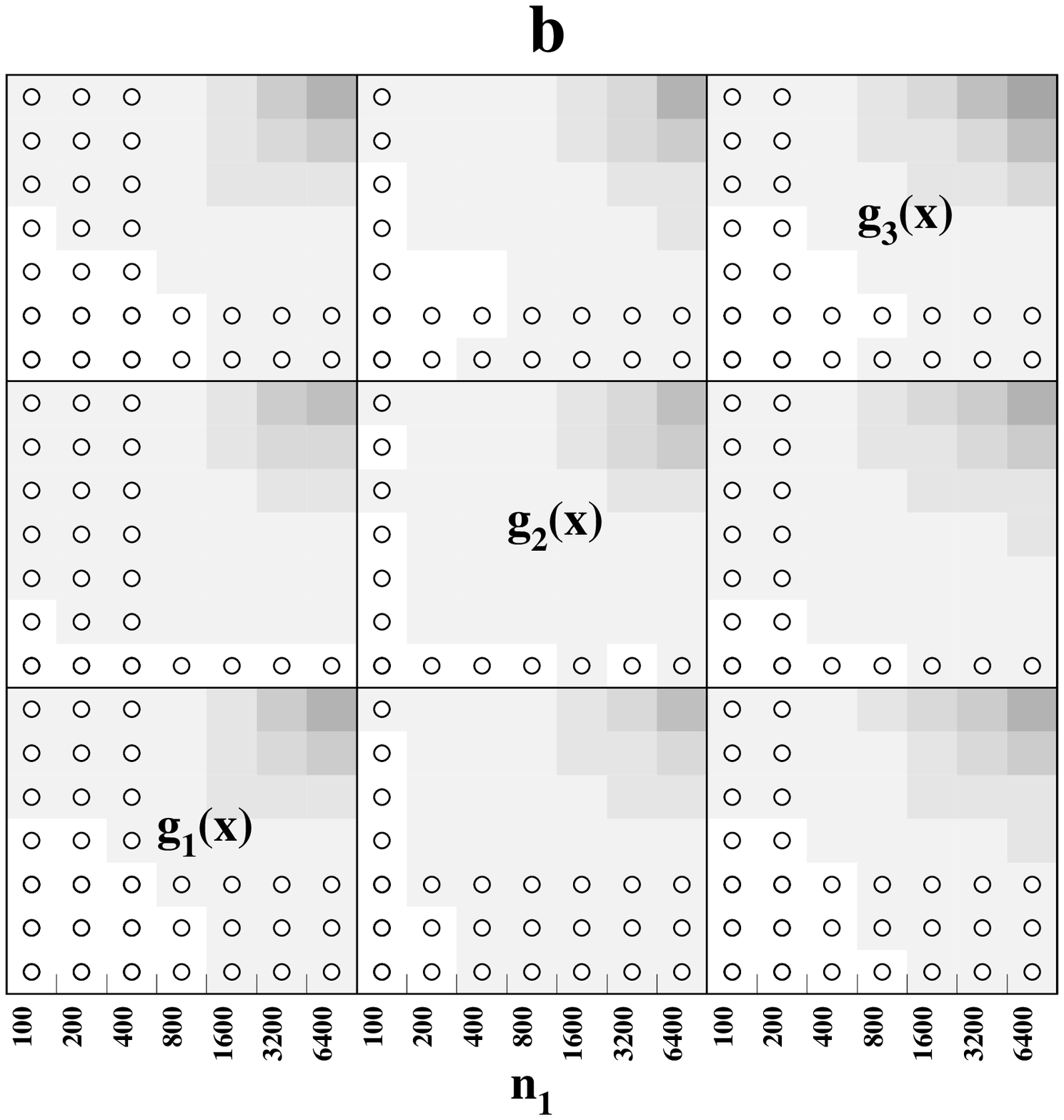} & \hspace*{-2.4 cm}\includegraphics[width=4.5in]{scalesizep.eps}
\end{array}$
\end{center}
\vspace *{-1cm}
\hspace{1cm}\parbox{19cm}{\caption{Powers of the chi-square test for the comparison of the histogram with normalized weights and the  histogram with unnormalized weights : (a) number of bins $m=5$, (b) number of bins $m=20$. Markers show regions with inappropriate number of events in the histograms for application of the test.}}
\end{figure}
\end{landscape}

\section{Conclusions}

In this study, a chi-square homogeneity test for the comparison of  histograms with normalized weights has been proposed. The test is a generalization of the classical homogeneity chi-square test. In addition, a test for histograms with unnormalized weights has also been developed. The proposed tests are very important tools in the application of the Monte Carlo method as well as in simulation studies of different phenomena. The evaluation of the sizes and powers of these tests was carried out numerically for histograms with different number of bins, events, and weight functions. The same investigation was carried out for the heuristic chi-square test that is currently being widely used. Comparison of the results showed the superiority of the new tests over the heuristic test.
The new tests can be used to fit the Monte Carlo data to the experimental data, compare the experimental data with the Monte Carlo data, compare two Monte Carlo data sets, and solve the unfolding problem by reweighting the events.


\begin{thebibliography}{00}
\bibitem{cramer}
H. Cramer, Mathematical methods of statistics, Princeton University Press, Princeton, 1999.
\bibitem{pearson}
K. Pearson, Phil. Mag. 50 (1900) 157-175.
\bibitem{fisher}
R.A. Fisher, J. Roy. Stat. Soc. 87 (1924) 442-450.
\bibitem{muon}
I. Abt et al., Eur. Phys. J. C50 (2007) 315-328.
\bibitem{weight}
A.M. Ferrenberg and R. H. Swendsen, Phys. Rev. Lett. 61 (1988) 2635-2638.
\bibitem{astro}
L.A. Pozdniakov, I.M. Sobol, R.A. Siuniaev, in: Soviet Scientific Reviews, Section E: Astrophysics and Space Physics Reviews, vol. 2, Harwood Academic, New
York, 1983, pp. 189-331.
\bibitem{hbook}
HBOOK-Statistical Analysis and Histogramming, Reference Manual, CERN,
Geneva, Switzerland, 1998.
\bibitem{paw}
PAW-Physics Analysis Workstation, User´s
 Guide, CERN, Geneva, Switzerland, 2001.
\bibitem{root1}
R. Brun et all, ROOT-An Object-Oriented Data Analysis Framework, Users Guide, CERN, Geneva, Switzerland, 2007.
\bibitem{Sobol}
I. Sobol, A Primer For The Monte Carlo Method, CRC Press, Boca Raton, Florida, 1994.
\bibitem{schmidt}
D.M. Schmidt, R.J. Morrison, M.S. Witherell, Nucl. Instr. Meth. A328 (1993) 547-552.
\bibitem{eberhard}
P. Eberhard, G. Lynch, D. Lambert, Nucl. Instr. Meth. A326 (1993) 574-580.
\bibitem{kortner}
O.K. Kortner, \v{C}. Zupan\v{c}i\v{c},  Nucl. Instr. Meth. A503 (2003) 625-648.
\bibitem{gagunashvili}
N. Gagunashvili, in:  Proceedings of the Conference on Statistical Problems in Particle Physics, Astrophysics and Cosmology, 12-15 September, 2005, Oxford, Imperial College Press, London, 2006, pp. 43-44.
\bibitem{gagunashvili1}
N. Gagunashvili, in: Proceedings of XI International  Workshop on Advanced Computing and Analysis Techniques in Physics Research, 23-27 April, 2007, Amsterdam, PoS(ACAT)060 http://pos.sissa.it//archive/conferences/050/060/ACAT\_060.pdf,  2007.
\bibitem{root}
http://root.cern.ch/root/htmldoc/TH1.html\#TH1:Chi2Test
\bibitem{zech}
G. Zech, Comparing statistical data to Monte Carlo simulation-parameter fitting and unfolding, DESY 95-113, June 1995, ISSN 0418-9833.
\bibitem{gagunashvili3}
N.D. Gagunashvili, Nucl. Instr. Meth. A596 (2008) 439-445.
\bibitem{brent}
R.P. Brent, Algorithms for Minimization without Derivatives, Chapter 4. Prentice-Hall, Englewood Cliffs, NJ, 1973.
\bibitem{cern}
CERN Program Library (D503), Reference Manual, CERN, Geneva, Switzerland, 1998.
\bibitem{moore}
D.S. Moore, G.P. McCabe, Introduction to the Practice of Statistics,
W.H. Freeman Publishing Company, New York, 2005.
\bibitem{cochran}
W.G. Cochran (1952), Ann. of Math. Stat. 23 (1952)  315-345.
\bibitem{breit}
G. Breit, E. Wigner, Capture of slow neutrons, Phys. Rev. 49 (1936) 519-531.
\bibitem{kendall}
M.G. Kendall, A.S. Stuart,  The Advanced Theory of Statistics, Vol.
2, Ch. 23, Griffin Publishing Company, London, 1973.
\end{thebibliography}
\end{document}